\documentclass[usenatbib]{mn2e}
\usepackage{multirow}
\usepackage{rotating}
\usepackage{colortbl}
\usepackage{color}
\usepackage{xcolor}
\usepackage[fleqn]{amsmath}
\usepackage{amssymb}
\usepackage{amsfonts}
\usepackage{verbatim}
\usepackage{scalefnt}
\usepackage{lscape}
\usepackage{booktabs}

\usepackage[bookmarks=False,bookmarksnumbered,colorlinks=true, citecolor=blue, linkcolor=cyan]{hyperref}

\newcommand{\apjl}{ApJ}% Astrophysical Journal, Letters
\newcommand{\apjs}{ApJS}% Astrophysical Journal, Supplement
\newcommand{\ao}{Appl.~Opt.}% Applied Optics
\newcommand{\aap}{A\&A}% Astronomy and Astrophysics
\newcommand{\beq}{\begin{equation}}
\newcommand{\eeq}{\end{equation}}
\newcommand{\mh}[1]{\textcolor{teal}{Melanie: #1}}

\title[Dual AGN in large-scale cosmological simulations]
{Large-scale dual AGN in large-scale cosmological hydrodynamical simulations}
\author[Puerto-S\'anchez et al.]{Clara Puerto-S\'anchez$^{1}$\thanks{E-mail: puertosanchezclara@gmail.com},
M\'elanie Habouzit$^{2,1}$, 
Marta Volonteri$^{3}$, Yueying Ni$^{4}$, 
\newauthor
Adi Foord$^{5}$, Daniel Anglés-Alcázar$^{6}$, Nianyi Chen$^{7}$, Paloma Guetzoyan$^{8}$, \newauthor
Romeel Dav\'e$^{8}$,
Tiziana Di Matteo$^{7}$, Yohan Dubois$^{3}$, Michael Koss$^{9}$, 
\newauthor
and Yetli Rosas-Guevara$^{10}$\\
$^1$ Max-Planck-Institut f\"ur Astronomie, K\"onigstuhl 17, D-69117 Heidelberg, Germany\\
$^2$ Department of Astronomy, University of Geneva, Chemin d'Ecogia, CH-1290 Versoix, Switzerland\\
$^{3}$ Institut d'Astrophysique de Paris, Sorbonne Universit\'es, CNRS, UMR 7095, 98 bis bd Arago, 75014 Paris, France\\
$^{4}$ Harvard-Smithsonian Center for Astrophysics, 60 Garden Street, Cambridge, MA 02138, USA\\
$^{5}$ Department of Physics, University of Maryland Baltimore County, 1000 Hilltop Cir, Baltimore, MD 21250, USA\\
$^{6}$ Department of Physics, University of Connecticut, 196 Auditorium Road, U-3046, Storrs, CT 06269-3046, USA\\
$^{7}$ McWilliams Center for Cosmology, Department of Physics, Carnegie Mellon University, Pittsburgh, PA 15213, USA\\
$^{8}$ Institute for Astronomy, Royal Observatory, University of Edinburgh, Edinburgh EH9 3HJ, UK\\ 
$^{9}$ Eureka Scientific, 2452 Delmer Street Suite 100, Oakland, CA 94602-3017, USA\\
$^{10}$ Donostia International Physics Centre (DIPC), Paseo Manuel de Lardizabal 4, 20018 Donostia-San Sebastian, Spain\\
}

\date{2024}

\begin{document}
\maketitle

\begin{abstract}
Detecting dual active galactic nuclei (DAGN) in observations and understanding theoretically which massive black holes (MBHs) compose them and in which galactic and large-scale environment they reside are becoming increasingly important questions as we enter the multi-messenger era of MBH astronomy.
This paper presents the abundance and properties of DAGN produced in nine large-scale cosmological hydrodynamical simulations. We focus on DAGN powered by AGN with $L_{\rm bol}\geqslant 10^{43}\, \rm erg/s$ and belonging to distinct galaxies, i.e. pairs that can be characterised with current and near-future electromagnetic observations.
We find that the number density of DAGN separated by a few to 30 proper kpc varies from $10^{-8}$ (or none) to $10^{-3} \, \rm comoving\, Mpc^{3}$ in the redshift range $z=0-7$. At a given redshift, the densities of the DAGN numbers vary by up to two orders of magnitude from one simulation to another.
However, for all simulations, the DAGN peak is in the range $z=1-3$, right before the peak of cosmic star formation or cosmic AGN activity.
The corresponding fractions of DAGN (with respect to the total number of AGN) range from 0 to $6\%$. We find that simulations could produce too few DAGN at $z=0$ (or merge pairs too quickly) compared to current observational constraints while being consistent with preliminary constraints at high redshift ($z\sim3$).
Next-generation observatories (e.g., AXIS) will be of paramount importance to detect DAGN across cosmic times. We predict the detectability of DAGN with future X-ray telescopes and discuss DAGN as progenitors for future LISA gravitational wave detections.

\end{abstract}


\section{Introduction}
The coming multi-messenger era of massive black holes (MBHs) will be pivotal to detecting MBHs closer to the redshift of their formation and merging MBHs (independently of whether they accrete or not). The James Webb Space Telescope, Euclid, Roman, and Athena telescopes (and NASA AXIS and Lynx concept missions) will significantly increase our ability to detect the electromagnetic emission of lower-mass and/or higher-redshift MBHs than is currently possible, while LISA will detect the gravitational waves from coalescing MBHs with $10^{4-7}\, \rm M_{\odot}$ at any redshift \citep[][for a review]{2023LRR....26....2A}. These new observatories will allow us to make headway in understanding the respective contributions of gas accretion and MBH mergers to MBH assembly across cosmic times and the connection between galaxy mergers and AGN fuelling. 
Understanding the dual AGN (DAGN) population theoretically and identifying and confirming them in observations will become increasingly important in the next few years. 
DAGN are defined as two accreting MBHs separated by a distance from sub-parsecs to tens of kiloparsecs; they can belong to the same galaxy or two close or interacting galaxies. 

In the hierarchical paradigm of galaxy formation,
we expect central MBHs to coalesce after the merger of their host galaxies. To do so, MBHs must cross an impressive range of scales, from tens of kpc galaxy separation to their coalescence \citep{1980Natur.287..307B}.
MBHs start by losing orbital energy and angular momentum via gravitational drag from background gas, stars, dark matter ({\it dynamical friction} phase, kpc scale), and sink to the centre of the remnant galaxy.
On $\lesssim$ pc scales, the MBHs form a gravitationally-bound binary and evolve further via interactions with gas and individual stars ({\it binary hardening} phase). The binary then interacts with its {\it circumbinary disc} on the ${\sim}10^{-3}$ pc scales (if gas is available), before entering the gravitational wave (GW) regime (${\leqslant}\, 10^{-5}\,\rm pc$ scale). 
These phases imply {\it delay times} between galaxy mergers and MBH coalescences, up to a Gyr timescale. While LISA will identify MBH mergers, we are currently mostly limited to identifying DAGN of kpc separation or more with electromagnetic observatories, and we thus dedicate our study to these systems.
We note, however, the existence of confirmed DAGN separated by hundreds of pc in observations \citep[][]{2023ApJ...942L..24K,2015ApJ...813..103M},  and that searches for pc or even sub-pc scale binaries are possible and rely on observations with the Very Long Baseline Array \citep[VLBA, e.g.][]{2016ApJ...826..106G}, specific emission lines, periodic light curves or radial velocity shifts of broad emission lines, or variability. 
At the time of writing, about a thousand kpc-scale separation DAGN candidates (including a handful of triple AGN systems) have been identified in the redshift range $z=0-3$ \citep[summarized in][and Fig.~\ref{fig:overview} in this paper]{2022ApJ...925..162C}. Less than 50 of those have been confirmed. A broad range of techniques are employed to identify these systems: optical spectroscopy in the local Universe with SDSS \citep[e.g.,][]{2013ApJ...769...95B} or HST at higher redshift \citep{2022ApJ...925..162C}, radio detections \citep[e.g., with VLBI][]{} with two bands to ensure that the spectra are flat and that the system is not an outflow, and X-ray detections \citep[e.g.,][]{2023MNRAS.519.5149D,2021ApJ...907...71F,2020ApJ...892...29F,2015ApJ...806..219C,2011ApJ...737L..19C}. 
Recent methods include varstrometry (from Gaia data) and HST follow-ups to confirm multiple sources \citep[and not lensed quasars,][]{2022ApJ...925..162C}, and Gaia multi-peak method coupled again to HST follow-ups \citep[][below 7 kpc separation]{2022NatAs...6.1185M,2023arXiv230507396M,2023A&A...671L...4C} or MUSE adaptive-optics spectroscopy \citep{2024A&A...690A..57S}.

Among the recent JWST discoveries, \citet{2023A&A...679A..89P} confirm the presence of an obscured AGN in one of the close clump companions of LBQS 0302-0019 quasar at $z\sim 3.3$, with a separation of $20 \, \rm kpc$, and \citet{2024arXiv240308098I} re-analyze the dual J0749+2255 quasars at $z=2.17$. About 30 DAGN candidates are identified in COSMOS-Web ($z=0-5$) and presented in \citet{2024arXiv240514980L}. The discoveries of several DAGN and a triple AGN in \citet{2023arXiv231003067P} in the redshift range $z\sim 3-3.7$ could hint towards a large fraction of these systems among the AGN population at high redshift.
In the near future, we expect several telescopes to substantially increase the number of detected DAGN, such as Euclid and Roman \citep[as discussed recently in Roman CSS White Papers,][]{2023arXiv230614990H,2023arXiv230615527S} or the Athena, AXIS, and LynX X-ray concept missions \citep[e.g.,][]{2023arXiv231100780R,2023arXiv231107664F}. In this paper, we predict the number of DAGN detectable by these observatories. 

Over the last decade, the community has developed large-scale cosmological hydrodynamical simulations of $\geqslant \, 100^{3} \, \rm comoving \,Mpc^{3}$ volume and $\sim \,1$ kpc spatial resolution (e.g., Horizon-AGN, Illustris, SIMBA) 
to understand galaxy formation. These simulations have been successful in producing galaxies with properties that are in good agreement with those of observed galaxies \citep[e.g., diversity of morphologies, stellar mass function,][]{2015ARA&A..53...51S,2015MNRAS.446..521S,2015MNRAS.450.1937C,2016MNRAS.463.3948D,2017arXiv170302970P}. The simulations allow us to capture the co-evolution between galaxies and their MBHs (to some level), but remain limited by their mass and spatial resolutions to capture the dynamics of MBHs and their mergers \citep[][for a review]{2023arXiv230411541D}. In fact, a spatial resolution of ${<}\,10 \, \rm pc$ is needed to resolve the first dynamical friction phase of MBH coalescence \citep{2019MNRAS.486..101P,2017ApJ...840...53R,2015MNRAS.451.1868T}. As a result, in most large-scale cosmological simulations, MBH particles are systematically repositioned to their host galactic centre (which also often ensures that MBHs are located in the galaxy's largest gas reservoir) and numerically merged instantaneously after the merger of their host galaxies. We note the exception of Horizon-AGN in which the effect of dynamical friction from the gas is modelled with a drag force \citep{2016MNRAS.463.3948D}; several MBHs can co-exist within the same galaxies. 
Recently, a drag force from the gas and the dynamical friction from dark matter and stellar particles were modelled in Astrid in \citet{2022MNRAS.510..531C} \citep[see also][]{2015MNRAS.451.1868T}.
To compute MBH merger rates from large-scale simulations without MBH dynamics, 
studies have included post-processing delays \citep{2016MNRAS.463..870S,2020MNRAS.498.2219V,2020MNRAS.491.2301K}. 
The merger rates for different MBH formation models were investigated in semi-analytical models, including delays \citep{2022MNRAS.509.3488I,2007MNRAS.377.1711S} or not \citep{2019MNRAS.486.2336D,2018MNRAS.481.3278R}. All the different modelling of MBH and galaxy physics and delays associated with the dynamical friction, hardening, and circumbinary disk phases have led to predictions on MBH merger rates spanning several orders of magnitude. To avoid the complex problem of MBH dynamics and modelling of possible delays, we restrict ourselves to the analysis of DAGN produced in large-scale simulations which live in separate galaxies and only briefly discuss their possible subsequent mergers. 

In this paper, we present a statistical study of the populations of dual MBHs and DAGN produced by the Illustris, TNG, Horizon-AGN, EAGLE, SIMBA, BlueTides, and Astrid simulations. We focus on dual systems that compose most of today's detections in observations and that the large-scale simulations can capture, i.e. those with separation in the range of a few to $\rm  30\, proper\, kpc$ and located in distinct galaxies. 
Our paper offers a summary of DAGN in the field of large-scale cosmological hydrodynamical simulations. It complements existing papers dedicated to EAGLE \citep{2019MNRAS.483.2712R}, Horizon-AGN \citep{2022MNRAS.514..640V}, and Astrid \citep{10.1093/mnras/stad834}. We note that the above papers limited their analysis to $z\leqslant 3-4$, while we extended our analysis to $z=10$ to predict what future missions dedicated to the early Universe could detect. Our work also completes previous studies with detailed and higher-resolution zoom-in and/or merger simulations \citep[][and references therein]{2015MNRAS.447.2123C,2017MNRAS.469.4437C} and post-processing approaches \citep[e.g.,][]{2022MNRAS.514.2220C}.\\

The paper is organised as follows. Section 2 presents the large-scale cosmological simulations, computation of AGN luminosity, and main differences identified among the simulations; we present the number density and fraction of DAGN in Sections 3 and 4, and their detectability by future observatories in Section 5. Conclusions are presented in Section 6.

\begin{table*}
\caption{Parameters of the cosmological simulations: box side length, number of baryons and dark matter particles, particle masses, maximum proper softening length ($\epsilon_{\rm DM}$), inter-particle spacing of gas particles at the star formation threshold ($\Delta x_{\rm gas,SF}$), initial MBH mass. 
}
\label{tab:resolution}
\begin{tabular}{lccccccccc}
\hline
                          & \textbf{Illustris} & \textbf{TNG50} & \textbf{TNG100} & \textbf{TNG300} & \textbf{HAGN}  & \textbf{EAGLE} & \textbf{SIMBA}  & \textbf{BlueTides} & \textbf{Astrid}\\ \hline
Box side {[}cMpc{]}       & 106.5   &      51.7     & 110.7           & 302.6           & 142            & 100            & 147.1      & 573.9 & 369.1   \\
$N_{\rm baryon}$                 & $1820^3$      &     $2160^3$  & $1820^3$          & $2500^3$          & -              & $1504^3$         & $1024^3$     & $7040^3$ & $5500^3$   \\
$N_{\rm DM}$                     & $1820^3$    &  $2160^3$       & $1820^3$          & $2500^3$          & $1024^3$         & $1504^3$         & $1024^3$    & $7040^3$ &   $5500^3$   \\
$M_{\rm baryon}$ {[}$\rm M_\odot${]}   & $1.3 \times 10^6$  & $8.5 \times 10^4$   & $1.4 \times 10^6$  & $1.1 \times 10^7$  & -              & $1.8 \times 10^6$ & $1.8 \times 10^7$ & $3.4 \times 10^6$ & $1.9 \times 10^6$ \\
$M_{\rm DM}$ {[}$\rm M_\odot${]}       & $6.3 \times 10^6$ &  $4.5 \times 10^5$  & $7.5 \times 10^6$  & $5.9 \times 10^7$  & $8.0 \times 10^7$ & $9.7 \times 10^6$ & $9.6 \times 10^7$ & $1.7 \times 10^7$ & $9.9 \times 10^6$ \\
$\epsilon_{\rm DM}$ {[}pkpc{]}     & 0.7    & 0.29           & 0.74            & 1.5             & -              & 0.7            & 0.74      & 2.2 & 2.2    \\
$\Delta x_{\rm gas,SF}$ {[}ckpc{]} & 0.7   &  0.28           & 0.7             & 1.4             & 1.0            & 1.1            & 0.53     & - &    -   \\ 
$M_{\rm MBH, \, seed}$ {[}$\rm M_\odot${]} & $1.4\times10^5$ & $1.2\times10^6$ & $1.2\times10^6$ & $1.2\times10^6$ & $10^5$ & $1.5\times10^5$ & $1.5\times10^4$ & $7.2\times10^5$ & $4.4 \times${[}$10^4$, $10^5${]} \\\hline
\end{tabular}
\end{table*}

\section{Methodology}
\subsection{Cosmological simulations}
\label{sec:methodology}
In this work, we use the following nine large-scale cosmological hydrodynamical simulations: Illustris, TNG50, TNG100, TNG300 (larger volume and lower resolution compared to TNG50 and TNG100), Horizon-AGN, EAGLE, SIMBA, BlueTides, and Astrid. 
All these simulations have volumes ranging from $50^{3}\, \rm cMpc^{3}$ to $\sim$$570^{3}\, \rm cMpc^{3}$, dark-matter mass resolutions of $\sim$$5\times 10^{5}-8\times 10^{7}\, \rm M_{\odot}$, and spatial resolutions of 1-2 ckpc. The parameters are presented in Table~\ref{tab:resolution}. All simulations run down to $z=0$, except BlueTides (snapshots are available for $z\geqslant 6.5$) and Astrid ($z\geqslant 3$).
The simulations model the evolution of the dark and baryonic matter contents in an expanding space-time, with different cosmologies (all consistent with WMAP or Planck). They capture the highly non-linear processes involved in the evolution of galaxies and MBHs, spanning kpc to Mpc scales.
For physical processes taking place at small scales below the galactic scale, they rely on subgrid modelling, e.g., for star formation, SN feedback, MBH formation, growth, and feedback. The subgrid models vary from simulation to simulation \citep[][for a summary]{2020arXiv200610094H}. More detailed descriptions of the simulations and their MBH modelling can be found in \citet{2014MNRAS.445..175G,2014MNRAS.444.1518V} for Illustris, \citet{2017arXiv170302970P,2018MNRAS.479.4056W} for TNG, \citet{2016MNRAS.463.3948D,2016MNRAS.460.2979V} for Horizon-AGN, \citet{2015MNRAS.446..521S,2016MNRAS.462..190R,2018MNRAS.481.3118M,2017MNRAS.468.3395M} for EAGLE, \citet{2019MNRAS.486.2827D,2019MNRAS.487.5764T,2020arXiv201011225T} for SIMBA, \citet{2016MNRAS.455.2778F} for BlueTides, and \citet{2022MNRAS.513..670N,2022MNRAS.512.3703B} for Astrid. \\

Cosmological simulations form MBHs in galaxies when certain seeding conditions are met.  MBHs form in all halos (e.g., when $M_{\rm h}\geqslant 1.5, \, 7.1, \,7.2$ or $7.4\times 10^{10}\, \rm M_{\odot}$ for EAGLE, Illustris, BlueTides and TNG, respectively), galaxies with $M_{\star}\geqslant 10^{9.5}\, \rm M_{\odot}$ (SIMBA), a combination of both with $M_{\rm h}\geqslant 7.4\times 10^{9}\, \rm M_{\odot}$ and $M_{\star}\geqslant 3\times 10^{6}\, \rm M_{\odot}$ for Astrid, or based on local properties in Horizon-AGN. 
The initial mass of the MBH is fixed in most simulations and varies from $10^{4}$ to $10^{6}\, \rm M_{\odot}$ between simulations. Unlike the other simulations presented here, in Astrid the initial mass of MBHs is drawn from a power-law distribution that covers the range $M_{\rm seed}=4.4\times 10^{4}- 4.4\times 10^{5} \, \rm M_{\odot}$.

MBHs accrete gas based on the Bondi-Hoyle model or a modified version of it taking into account the gas angular momentum (EAGLE) or distinguishing between the accretion of cold gas ($T<10^{5}\, \rm K$) driven by gravitational torques \citep{2017MNRAS.464.2840A} and Bondi accretion of hot gas ($T>10^{5}\, \rm K$) for SIMBA.

The Bondi-Hoyle model is described by: 
\begin{equation}
    \dot{M}_{\rm MBH} = \frac{4\pi\rho G^2M_{\rm MBH}^2}{(c_{\rm s}^2+v_{\rm rel}^2)^{3/2}},
\end{equation}
with $\rho$ and $c_s$ being the gas density and sound speed, $G$ the gravitational constant, $M_{\rm BH}$ the mass of the MBH and $v_{\rm rel}$ the relative velocity between the gas and the MBH. Accretion rates are capped at the Eddington accretion limit ($\dot{M}_{\rm Edd}$), except in Astrid whose MBHs can accrete for short periods of time up to twice $\dot{M}_{\rm Edd}$.

Large-scale simulations do not have the resolution to capture BH dynamics, but some have a subgrid implementation for dynamical friction.

When moving, BHs create an overdensity of material (gas, stars, and dark matter) behind them, which in return drag and decelerate them and eventually help them to sink to the center of their host galaxies. Horizon-AGN and Astrid use a drag force to model the dynamical friction from the gas \citep{2016MNRAS.463.3948D,2022MNRAS.513..670N}, and Astrid has an additional model for dynamical friction from dark matter and stellar particles \citep{2022MNRAS.510..531C}. 
These implementations prevent regularly replacing BHs to their galaxy potential well and allow for several BHs per galaxies (as a consequence of galaxy mergers). The other simulations re-position BHs to the location with the lowest gravitational potential \citep{2022MNRAS.516..167B}. For example, this minimum potential is searched in a region containing 1000 mass resolution elements in TNG. To avoid spurious moves of the BHs (e.g., galaxies loosing their BHs, or BHs jumping from their satellite galaxies to close central galaxies), some simulations like EAGLE have additional criteria such as a smaller distance of three gravitational softening lengths between the BH and its future location as well as a relative velocity smaller than 0.25 the speed of sound. As described in Section~\ref{sec:identification_dual_AGN}, we focus in this paper on dual AGN systems whose AGN are located in distinct galaxies. This allows a common ground to compare the simulations with different models of dynamics and repositioning. The main effect of the different schemes on the current study is the possible enhancement of DAGN by making AGN more frequent in general. BHs that are re-positioned into the gravitational potential well are more likely to be located in a gas reservoir and to accrete more gas. However, artificially making MBHs merge very quickly can have the opposite effect (decreasing the number of DAGN), since the repositioning makes MBHs "jump" quickly to the center of the main galaxy/halo where the MBHs can quickly merge rather than spending more time orbiting as DAGN.

The spatial and mass resolutions of the simulations, listed in Table~\ref{tab:resolution} (with softening lengths larger than $0.5\, \rm pkpc$), also limit their ability to elucidate the role of gas dynamics in triggering AGN phases in dual or binary systems. At least an order of magnitude higher resolution is necessary to accurately capture the effects of small-scale loss of angular momentum, gas inflow, and outflow. For instance, 
\citet{2015MNRAS.447.2123C,2017MNRAS.469.4437C} employ galaxy merger simulations with dark matter particle softening lengths of 20-30 pc and MBH softening lengths of 5 pc to study gas dynamics on small scales. Their analysis revealed that the specific angular momentum around the secondary MBH does not change until the second pericentric passage, when the separation drops from tens of kpc to tens of pc, a regime that we do not address here with dual AGN in separate galaxies.

Finally in this paper, we do not explicitly distinguish between radiatively efficient and 
inefficient AGN. We assume that MBHs are all radiatively efficient and compute their bolometric luminosity as:
\begin{eqnarray}
L_{\rm bol}=\epsilon_{\rm r} \dot{M}_{\rm MBH} c^{2}.
\label{eqn:method1}
\end{eqnarray}
We use the same radiative efficiency of $\epsilon_{\rm r}=0.1$ for all the simulations, which provides a better agreement with current observational constraints. 
We adopt the model of \citet{2020MNRAS.495.3252S} to convert the bolometric luminosity to the hard X-ray band (2-10 keV):
\begin{eqnarray}
\frac{L_{\rm bol}}{L_{\rm x}} = \alpha_{1} \left(\frac{L_{\rm bol}}{10^{10}\, {\rm L_{\odot}}} \right)^{\beta_{1}} + \alpha_{2} \left(\frac{L_{\rm bol}}{10^{10}\, {\rm L_{\odot}}} \right)^{\beta_{2}},
\label{eq:xray}
\end{eqnarray}
with $\alpha_{1}, \alpha_{2}=12.60, 4.073$ and $\beta_{1}, \beta_{2}=0.278, -0.026$.

\begin{figure*}
    \centering
    \includegraphics[scale=0.7]{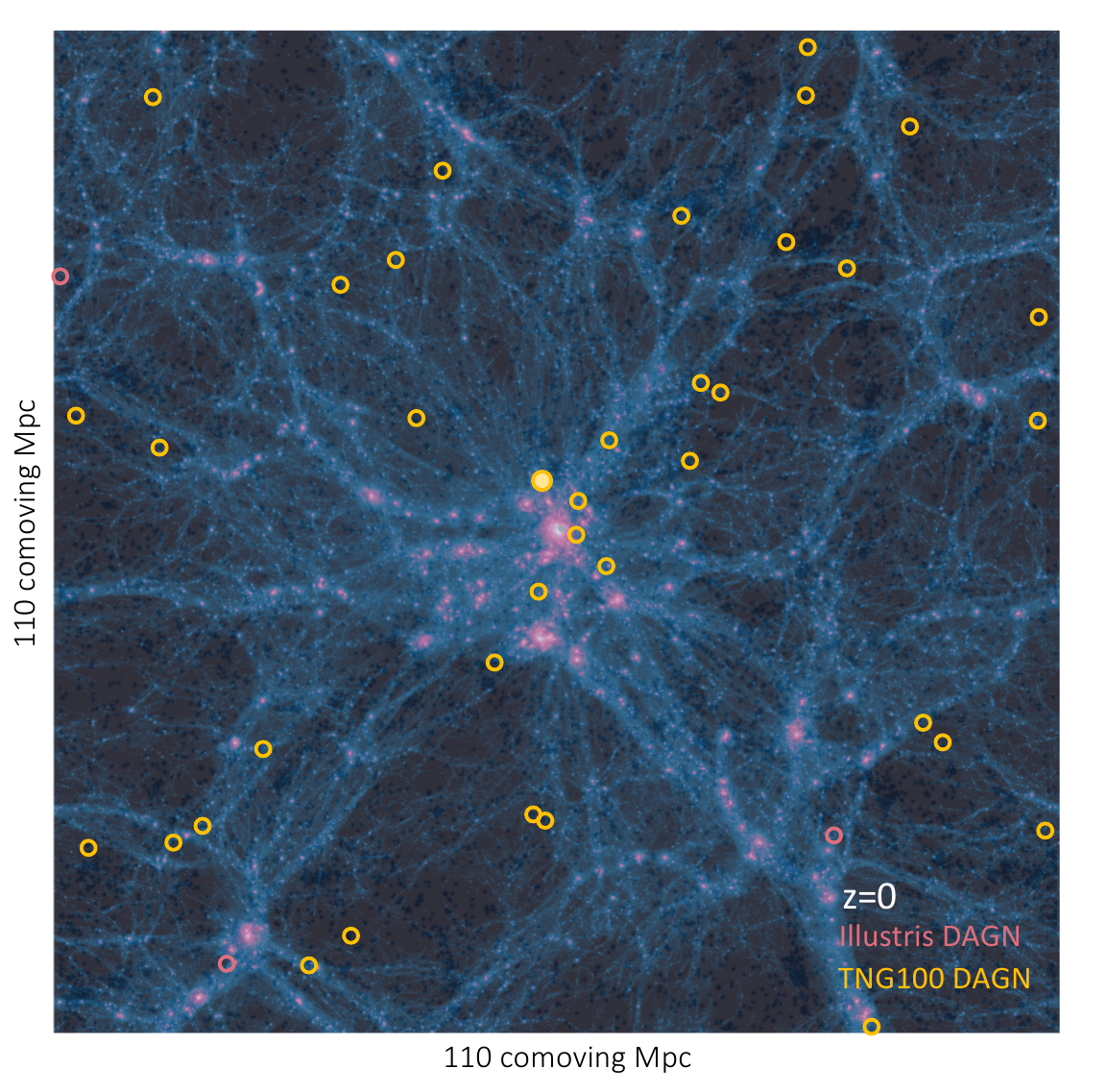}
    \caption{Map of Illustris dark matter density combined with stellar composite images with a depth of 10 Mpc; a light scatter plot shows the full population of galaxies.
    The location of the DAGN identified at $z=0$ in Illustris and TNG100 is shown with pink and yellow symbols, respectively. The two simulations share the same initial conditions and, therefore, have similar large-scale structures. The differences in the subgrid physics have a strong impact on the number density of DAGN: only 3 DAGN are produced in Illustris against 36 in TNG100; the latter additionally produces a triplet (filled yellow symbol) where DAGN are separated by 15, 23, 27 kpc.}
    \label{fig:map}
\end{figure*}

\subsection{Summary of the main differences in simulated MBH and AGN populations}
The modelling of subgrid galaxy and MBH physics varies from one simulation to another and leads to differences in the population of MBHs \citep{2021MNRAS.503.1940H} and AGN \citep{2021MNRAS.tmp.2905H} produced by the simulations. 
We briefly summarise below the important aspects identified in the above papers that are relevant to the present paper.

Most of the cosmological simulations have a tighter $M_{\rm BH}-M_{\star}$ relation than the current observations in the local and low-redshift Universe \citep[e.g.,][]{2015arXiv150806274R,2018ApJ...869..113D,2019MNRAS.487.3404B}. At fixed total galaxy stellar mass, the scatter in $M_{\rm MBH}$  is smaller in simulations than in observations. In particular, the region of the $M_{\rm MBH}-M_{\star}$ diagram corresponding to observed broad-line AGN (i.e., MBHs of $M_{\rm MBH}=10^{5.5}-10^{7.5}\, \rm M_{\odot}$ located in galaxies with $M_{\star}=10^{10}-10^{11}\, \rm M_{\odot}$) is not well reproduced in many simulations. The distribution of  MBH initial masses used in Astrid instead of a fixed mass produces a larger scatter in $M_{\rm MBH}$ in low-mass galaxies compared to the other simulations \citep{2022MNRAS.513..670N}.
Additionally, there is no consensus on the normalisation and shape of the mean $M_{\rm MBH}-M_{\star}$ relation in the simulations and how these aspects evolve with redshift. 

The properties of AGN are not calibrated in these large-scale simulations, so they are true predictions that can be judged against observational constraints.
While the simulations' AGN luminosity function broadly agree at $z=0$, discrepancies arise and increase at higher redshift. On average, all these simulations overproduce the number of AGN with $L_{\rm bol}\leqslant 10^{45}\, \rm erg/s$ at $z\sim 4$ with respect to observational constraints \citep[e.g.,][]{Hop_bol_2007,2010MNRAS.401.2531A,2015MNRAS.453.1946G,2015ApJ...802...89B,2015MNRAS.453.1946G}, but are in good agreement for more luminous objects. 
Among all the simulations, EAGLE has the lowest luminosity function at any redshift.
We note that the AGN population in the recent Astrid does not show such an excess of AGN \citep[][their Fig.~3]{2022MNRAS.513..670N}.
In general, the agreement on the faint end of the simulations improves when all AGN embedded in low-mass galaxies are removed with $M_{\star}\leqslant 10^{10}\, \rm M_{\odot}$ \citep{2021MNRAS.tmp.2905H}. This indicates that simulations may produce too many AGN in low-mass galaxies due to too massive MBHs, too efficient gas accretion, and/or too weak SN feedback \citep{habouzit2017blossoms,2017MNRAS.472L.109A}, as demonstrated in \citet{2022MNRAS.514.4912H}.
Astrid is the simulation that produces the lowest normalisation of the AGN luminosity function for $L_{\rm bol}\geqslant 10^{43}\, \rm erg/s$, at $z\geqslant 3$ (current snapshots available). At lower redshift, EAGLE produces the lowest number of AGN and is in good agreement with the observations mentioned above at $z\sim4$ for the faint end of the luminosity function, but falls short for the bright end and may not produce enough AGN with $L_{\rm bol}\geqslant 10^{45}\, \rm erg/s$. 

These differences with current observations must be considered when interpreting the predictions on DAGN presented in this paper. 

\subsection{Identification of dual AGN}
\label{sec:identification_dual_AGN}
This paper focusses on dual MBHs with three-dimensional separations of $d\leqslant 30 \, \rm pkpc$ and belonging to distinct galaxies. With the latter criterion, we avoid entering a regime where MBH dynamics needs to be captured and implemented (which is not the case in most simulations, see Section~\ref{sec:methodology}) to follow the evolution of MBHs from the merger of their galaxies to their coalescence. 
In this paper, we do not restrict ourselves to massive MBHs, but instead consider all the MBHs present in each simulation. We note that, for example, the authors of \citet{10.1093/mnras/stad834} (Astrid) have limited their analysis to MBHs with $M_{\rm MBH}\geqslant 10^{7}\, \rm M_{\odot}$ 
to ensure to capture with sufficient accuracy of the dynamics of MBHs (which is modelled in that simulation) belonging to the same galaxies. 
Finally, these 9 simulations can be split into two categories: i) simulations which allow for several MBHs to co-exist in the same galaxies (this is the case, e.g. when MBHs are not repositioned immediately to galaxy centres (or repositioning is done within a few smoothing lengths below the galaxy radius), or if MBH dynamics is captured to some level), and ii) simulations for which the majority of galaxies (except a few) only have one central MBH. For case i), we consider only the most massive MBH present in the galaxies (e.g., Astrid, SIMBA), and for a couple of galaxies with several MBHs in case ii), we sum up the masses of the MBHs (e.g., Illustris, TNGs). For Horizon-AGN, we employ the methodology described in \citet{2022MNRAS.514..640V} and only work on snapshots corresponding to $z\leqslant6$, and additionally enforce that only the most massive MBH is kept per galaxy.

\begin{figure*}
    \centering
    \includegraphics[scale=0.58]{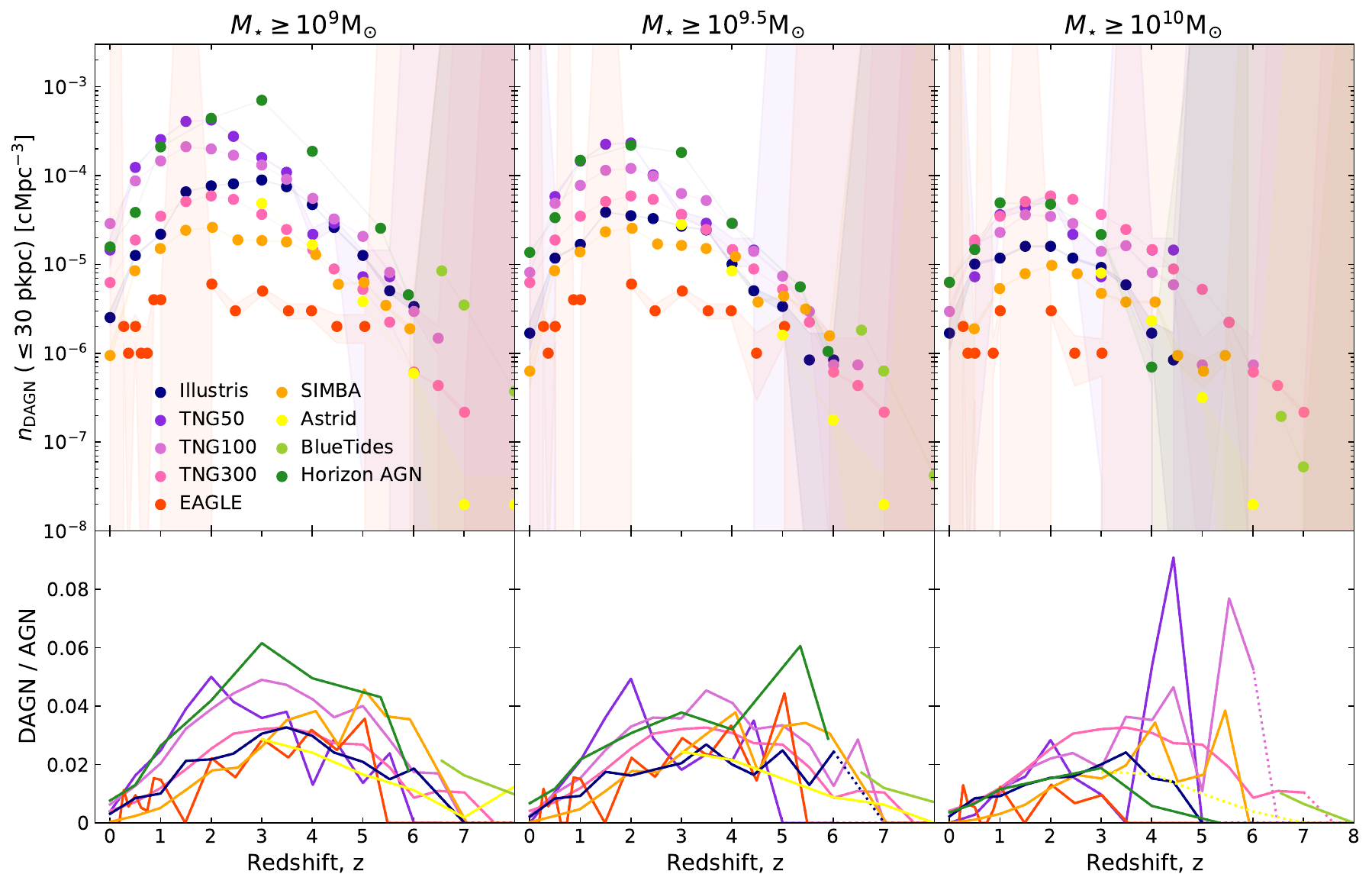}
    \caption{\textit{Top panels:} Number density of DAGN as a function of redshift. DAGN are defined as AGN with $L_{\rm bol} \geqslant 10^{43}$ erg/s separated by $\leqslant 30$ pkpc. Shaded regions indicate Poisson errors. We apply different mass cuts: only galaxies with $M_\star \geqslant 10^9, \,10^{9.5},\, 10^{10} \, \rm M_\odot$ are considered on the \textit{left}, \textit{central}, \textit{right panels}, respectively. 
    \textit{Bottom panels:} Fraction of DAGN, defined as the number of DAGN divided by the total number of AGN with the same luminosity cut. We only display the fractions when at least 10 AGN are identified at a given redshift. When there are less than 10 AGN identified, we show the fraction as a dotted line.}
    \label{fig:ndensity_fraction_z}
\end{figure*}

\section{Number density and fraction of dual AGN}

To illustrate how the subgrid physics implemented in simulations can dramatically impact the number of DAGN (and the properties of the MBHs and galaxies that power them), we present in Fig.~\ref{fig:map} the DAGN identified in Illustris and TNG100 at $z=0$. MBHs are considered AGN when $L_{\rm bol} \geqslant 10^{43} \, \rm erg/s$, and DAGN if the three-dimensional distance between them is $d \leqslant 30$ pkpc.
Illustris produces 3 DAGN at that time.
Because Illustris and TNG100 share the same initial conditions and, as such, the same large-scale structures and positions of galaxies, we also highlight the TNG100 DAGN: 36 DAGN, including a triple AGN system (filled symbol).
In the following, we quantify the different DAGN populations produced in cosmological simulations.

\subsection{Number density of dual AGN}
\subsubsection{Large differences among simulations}

In Fig.~\ref{fig:ndensity_fraction_z}, DAGN are identified with 
$L_{\rm bol} \geqslant 10^{43} \, \rm erg/s$ 
and a three-dimensional separation of 
$d \leqslant 30$ pkpc. 
A given AGN can be part of several DAGN: triple systems are counted as two DAGN in this section and explained further in Section~\ref{sec:multiple_AGN}.
In this paper, we mostly restrict our analysis to galaxies with $M_{\star}\geqslant 10^{9}\, \rm M_{\odot}$, if not indicated otherwise. Although we use a limit of $d \leqslant 30$ pkpc, only a small fraction of DAGN are found with separation of $d \leqslant 5$ pkpc (see Fig.~\ref{fig:hist_dual_d_60} and Section~\ref{sec:separation}). This happens because we restrict ourselves to DAGN in separate galaxies.

We find that the number density of DAGN varies from $10^{-7}$ to $10^{-3}\, \rm cMpc^{-3}$ between simulations and in the redshift range $z=0-7$ (top panels in Fig.~\ref{fig:ndensity_fraction_z}). No DAGN are identified at higher redshifts, except in the much larger simulated volume of BlueTides. 
EAGLE is the simulation that produces the lowest number of DAGN. As shown in Fig.~\ref{fig:DAGN_follow_AGN} below, it is also the simulation that produces the fewest AGN (i.e., the lower normalisation of the AGN luminosity function) and the overall faintest population of AGN \citep{2021MNRAS.tmp.2905H} at $z\leqslant 3$ due to a combination of factors, including a modified Bondi accretion model accounting for the angular momentum of the material falling into MBHs \citep{2015MNRAS.454.1038R,2016MNRAS.462..190R}: EAGLE MHBs thus have a harder time to accrete gas efficiently. For that simulation, the mass accretion rates are the instantaneous ones and can vary significantly from one snapshot to another one \citep[see Fig.~1 in][]{2019MNRAS.483.2712R} and most intense periods of AGN activity could be missed. In any case, the absence of DAGN in EAGLE is driven by the absence of AGN rather than a low MBH clustering and, indeed, dual MBHs are not rare; the simulation includes 160 dual MBHs at $z=0$.
Similarly to EAGLE, SIMBA and Astrid (at high redshift), which also have a relatively faint population of AGN, produce a number density of DAGN on the lower end of the full range covered by the simulations.

On the contrary, the large seeding mass employed in TNG and the weaker SN feedback in low-mass galaxies at lower redshift \citep{2018MNRAS.473.4077P} produce a large number of AGN in the TNG simulations compared to the other ones. We find the same results for Horizon-AGN, with the high DAGN number densities being driven by the large population of AGN with $L_{\rm bol}\leqslant 10^{45}\, \rm erg/s$ at $z\geqslant 2$ compared to the other simulations \citep{2021MNRAS.tmp.2905H}. In Horizon-AGN, the high number density of AGN and DAGN is partly due to an excess in the number density of galaxies compared to the current observational galaxy mass function \citep{2017MNRAS.467.4739K}.

As shown in Fig.~\ref{fig:DAGN_follow_AGN}, the number densities of DAGN follow those of AGN, with EAGLE, SIMBA, and Astrid producing the lowest number density of DAGN and the TNG50/TNG100 and Horizon-AGN the highest numbers.

Although the simulations show a large difference in the number densities of DAGN (up to 2 orders of magnitude, Fig.~\ref{fig:ndensity_fraction_z}), there is a consensus that the highest number of pairs is produced at cosmic noon, i.e. $z=1-3$. This is in good agreement with the number of DAGN and candidates identified in observations (see Fig.~\ref{fig:overview}). We find that DAGN peaks at slightly higher redshifts than the AGN number density and cosmic star formation rate. As further discussed in Section~\ref{sec:galaxies} and Fig.~\ref{fig:hist_Lbol}, beyond the local Universe, simulated DAGN tend to be on average brighter than the AGN population (when considering the same luminosity cut of $L_{\rm bol}\geqslant 10^{43}\, \rm erg/s$). As a result of AGN downsizing \citep[e.g.,][]{2004MNRAS.353.1035M,2014MNRAS.442.2304H}, the population of DAGN thus reaches its cosmic peak of activity before the AGN population (Fig.~\ref{fig:DAGN_follow_AGN}).

A summary table of the number of dual MBHs and DAGN is provided in Table~\ref{tab:dual_AGN_30} for $z=4,\, 2,\,0$.

\subsubsection{Number density for different stellar mass cuts}
In the left panel of Fig.~\ref{fig:ndensity_fraction_z} we consider all galaxies with a total stellar mass of $M_{\star}\geqslant 10^{9}\, \rm M_{\odot}$, galaxies with $M_{\star}\geqslant 10^{9.5}\, \rm M_{\odot}$ and $M_{\star}\geqslant 10^{10}\, \rm M_{\odot}$ in the middle and right panels, respectively. 
For most of the simulations (TNG100, TNG300, SIMBA, and EAGLE), the number density of DAGN and the shape of their evolution with redshift are not strongly impacted when only considering galaxies with $M_{\star}\geqslant 10^{9.5}\, \rm M_{\odot}$ (less than half of dex for the largest differences), indicating that the population of DAGN is not hosted in dwarf galaxies with $M_{\star}=10^{9}-10^{9.5}\, \rm M_{\odot}$ for the most part. 
In SIMBA, MBH seeding takes place in galaxies of $M_{\star} \geqslant 10^{9.5}\, \rm M_{\odot}$, which explains why the number density is almost\footnote{Some SIMBA galaxies with $M_{\star}< 10^{9.5}\, \rm M_{\odot}$ host a MBH while they should not according to SIMBA's seeding prescriptions \citep{2022MNRAS.514.4912H}; this can be explained by these galaxies forming a MBH when more massive and then loosing material, e.g. by being stripped by another more massive nearby galaxy.} identical in the left and middle panels.  
On the other hand, Illustris and TNG50 show a clear difference when introducing the $M_{\star}\geqslant 10^{9.5}\, \rm M_{\odot}$ mass cut (especially towards higher redshifts), which means that many of the DAGN in these simulations are linked to galaxies with smaller stellar masses. 
We develop this aspect in Section~\ref{sec:galaxies}. \\

\subsubsection{Number density at the highest redshifts}
The large volume of BlueTides allows us to investigate DAGN at the highest redshifts, up to $z=10$ (when the first DAGN is found). We find that BlueTides tends to produce more DAGN (per unit volume) than the extrapolation from lower redshifts of the other simulations when considering $M_{\star}\geqslant 10^{9}\, \rm M_{\odot}$. This is caused by the high seeding mass of $7.2\times 10^{5}\, \rm M_{\odot}$ and the efficient MBH growth (especially for $M_{\star}\geqslant 10^{9.5}\, \rm M_{\odot}$). 
Interestingly, we find that the number density of DAGN hosted by galaxies with $M_{\star}\geqslant 10^{10}\, \rm M_{\odot}$ is lower in BlueTides than in TNG300 (one of the other simulations with $\geqslant 300 \, \rm cMpc$ side length). 
Although we find the presence of a large population of MBHs with $M_{\rm MBH}\geqslant 10^{8}\, \rm M_{\odot}$ at $z\geqslant 7$ in $\geqslant 10^{10}\, \rm M_{\odot}$ galaxies (a population that cannot be present in the other smaller-volume simulations, even in TNG300 and Astrid) able to accrete with $L_{\rm bol}\geqslant 10^{44}\, \rm erg/s$, they are not proportionally involved in AGN pairs (see DAGN fractions in Fig.~\ref{fig:silverman2020}).\\

\subsubsection{Impact of simulation resolution}
We assess the impact of simulation resolution in Fig.~\ref{fig:ndensity_resolution} with the sibling TNG simulations (TNG50 being the highest-resolution simulation and TNG300 the lowest) whose subgrid models are identical \citep[not recalibrated for the different simulations, see][]{2018MNRAS.473.4077P}.
In TNG300, the less resolved gas content surrounding the MBHs leads to lower accretion rates and fewer AGN pairs with respect to TNG50 and TNG100. As illustrated in Fig.~\ref{fig:ndensity_resolution}, gas resolution strongly impacts the number of DAGN whose number density can vary by more than a dex with the different TNG resolutions.

With its higher resolution, TNG50 resolved galaxies down to $M_{\star}\geqslant 10^{7}\, \rm M_{\odot}$ (shown as a dashed line in Fig.~\ref{fig:ndensity_resolution}), and thus captured the assembly of dwarf and low-mass galaxies with $M_{\star}\sim 10^{9}\, \rm M_{\odot}$ (i.e., captures their growth on two orders of magnitude in stellar mass).
We add the number density of TNG50 DAGN when considering $M_{\star}\geqslant 10^{7}\, \rm M_{\odot}$ with open symbols in Fig.~\ref{fig:ndensity_resolution}. The inclusion of galaxies with $M_{\star}=10^{7}-10^{9}\, \rm M_{\odot}$ significantly increases the predicted number density of DAGN. It also demonstrates that number densities of AGN pairs with $d\leqslant 30 \, \rm pkpc$ in distinct galaxies that are derived from large-scale cosmological simulations only resolving galaxies with $M_{\star}\geqslant 10^{9}\, \rm M_{\odot}$ can be underestimated, and so are predictions with lower resolution (e.g., TNG300).

\subsection{Fraction of dual AGN and comparison with observational constraints}
%\textcolor{red}{Definition of fraction}
In the bottom panels of Fig.~\ref{fig:ndensity_fraction_z}, we show the fraction of DAGN for the different simulations, which quantifies out of all the AGN in the simulation how many pairs are found. The fraction is thus defined as ``DAGN / AGN'' where the numerator is the number of DAGN pairs in a given simulation and the denominator is the total number of AGN in that simulation.
The fractions range from $\sim 0.01$ to $\sim 0.06$ 
in the redshift range $z \sim 2-6$ (i.e., 1 to $6\%$ of AGN with $L_{\rm bol}\geqslant 10^{43}\, \rm erg/s$ belongs to DAGN). \\

Although the evolution of the number densities of DAGN with redshift follows the evolution of the AGN (Fig.~\ref{fig:DAGN_follow_AGN}), we find that the DAGN fractions do not necessarily exactly follow the number densities of DAGN: a simulation can have the highest normalisation of the number density while not having the highest DAGN fraction compared to the other simulations, for example.  This is because the difference between the DAGN and AGN number densities varies from one simulation to another.
For example, TNG100 produces a higher fraction of DAGN than TNG50 at $z\geqslant 2.5$, while the TNG100 DAGN number density of TNG100 is smaller than TNG50 up to $z\geqslant 3.5$.  This is due to a smaller increase in the TNG50 DAGN with respect to the increase in the TNG50 AGN over time, in the redshift range $\sim 2.5-3.5$. \\

Our findings on the DAGN fractions are consistent with previous work with the same simulations, which, for some of them, employed different selection criteria \citep[][for a compilation]{2019NewAR..8601525D}. \citet{2019MNRAS.483.2712R} study DAGN with $f_{\rm Edd}\geqslant 0.01$ and $L_{\rm x,\, 2-10\, keV}\geqslant 10^{42}\, \rm erg/s$ with separation of $1-30 \, \rm pkpc$, and find that the DAGN fraction increases with redshift in the range $0-3\%$.
With Horizon-AGN, \citet{2022MNRAS.514..640V} find an increasing fraction of DAGN with redshift, ranging from 0.01 at $z=0$ to $0.063\%$ at $z=3$, and number density ranges from $10^{-5}$ to $10^{-3}\,\rm cMpc^{3}$.
\citet{10.1093/mnras/stad834} report a DAGN fraction (with $M_{\rm BH}\geqslant 10^{7}\, \rm M_{\odot}$) of $3\%$ with a separation below 30 kpc at $z=2$, most of which are within the same galaxies and for separations of $<5\, \rm kpc$.

As developed in Section~\ref{sec:study_dependencies}, the evolution of the DAGN fractions over time (both the amplitude and the shape with redshift) can depend on the DAGN selection criteria (AGN luminosity, MBH mass, and host stellar mass).

\subsubsection{Comparison with observational constraints}
In this section, we compare the fraction of DAGN with observational constraints. Depending on the characteristics of the observations and surveys, these constraints can translate in different definitions of the fraction to be computed from the simulations and can significantly impact the results. 
For example, the fraction can be written as ``number of AGN involved in DAGN pairs / total number of AGN'' or ``number of DAGN pairs / (total number of AGN - number of DAGN pairs)'' which can be used when AGN companion(s) are not part of the initial AGN sample and identified through new observations. In Fig.~\ref{fig:comp_obs}, we gather these 3 main definitions, and show several observational constraints with grey and black symbols (reproduced identically in all panels). 

We start by comparing with constraints in the low-redshift Universe.
\citet{2011ApJ...737..101L} study AGN pairs (and several multiple AGN) with a projected separation of $\leqslant 70 \, \rm kpc$ and in the redshift range $z=0.02-0.33$ (mostly $z=0.02-0.16$) from SDSS DR7. 
AGN are classified with emission lines and line ratios. They find a DAGN fraction of $1.85\%$ (after correction for SDSS spectroscopic incompleteness) with a projection separation below $30 \, \rm kpc$ and higher than $3.5\, \rm kpc$ (numbers extracted from their Fig.~11). 
Many of these DAGN are found in interacting galaxies in mergers and present tidal features ($60\%$). The fraction of DAGN has been found to vary quite significantly as a function of the sensitivity of the AGN identification \citep{2011ApJ...737..101L}. 
\citet{2012ApJ...746L..22K} study DAGN in the BAT survey, i.e., X-ray selected AGN for $z<0.05$, and combine data with e.g. SDSS to identify companions and derive a DAGN fraction. DAGN are selected as having $L_{\rm 2-10\, keV}>10^{42}\, \rm erg/s$ (while our limit in Fig.~\ref{fig:ndensity_fraction_z} is $L_{\rm bol}\geqslant 10^{43}\, \rm erg/s $), or a hard power-law spectrum, or a Fe K line, or rapid time variability, or on optical emission line diagnostics. These constraints are closer in definition to the ones we use in the middle and bottom rows in Fig.~\ref{fig:comp_obs}.
At low redshift, we find that the simulations tend to produce a lower number DAGN fraction than currently observed, even more so if we only consider DAGN in galaxies with at least $M_{\star}\geqslant 10^{9.5}, \, 10^{10}\, \rm M_{\odot}$. Still, it is important to mention here that error bars in the observations can be large and highly sensitive to how DAGN are defined and which exact definition of fraction is used.\\

We also show in Fig.~\ref{fig:comp_obs} the upper limit on the fraction of DAGN at $z=3$ derived in \citet{2023arXiv231202311S}. 
The authors analysed a sample of 64 X-ray sources from the COSMOS-Legacy Survey \citep{2016yCat..18190062C}, X-UDS \citep{2018yCat..22360048K}, AEGIS-XD \citep{2015yCat..22200010N} and CDFS \citep{2017ApJS..228....2L} with $L_{\rm 0.5-8\, keV}>10^{43}\, \rm erg/s$ and an off-axis angle smaller than $< 5'$ for $2.5 < z < 3.5$, in which no DAGN has been identified. 
Our limit of $L_{\rm bol}\geqslant 10^{43}\, \rm erg/s$ used in Fig.~\ref{fig:comp_obs} is less restrictive than the selection criterion used in the above observations. For that reason, we also show in Fig.~\ref{fig:ndensity_fraction_z_44} the DAGN fractions for the most luminous systems with $L_{\rm bol}\geqslant 10^{44}\, \rm erg/s$.
We find that all the simulations are consistent with the observational constraint at $z\sim3$ when considering $M_{\star}\geqslant 10^{9.5}\, \rm M_{\odot}$ or $\geqslant 10^{10}\, \rm M_{\odot}$. When considering smaller galaxies with $M_{\star}\geqslant 10^{9}\, \rm M_{\odot}$, we note that TNG100 and Horizon-AGN produce a higher fraction of DAGN with $L_{\rm bol}\geqslant 10^{43}\, \rm erg/s$ than in the observations at $z=3$ (see next paragraph), but similar fractions to the constraint when considering only brighter DAGN (Fig.~\ref{fig:ndensity_fraction_z_44}).
All considered, most of the simulations lie on the upper side of the constraint (while we have excluded DAGN living in the same galaxies). 
However, we cannot rule out any of the simulations with these results. Additional observational constraints are needed to disentangle simulations and constrain their modelling.

\begin{figure}
    \centering
    \includegraphics[scale=0.41]{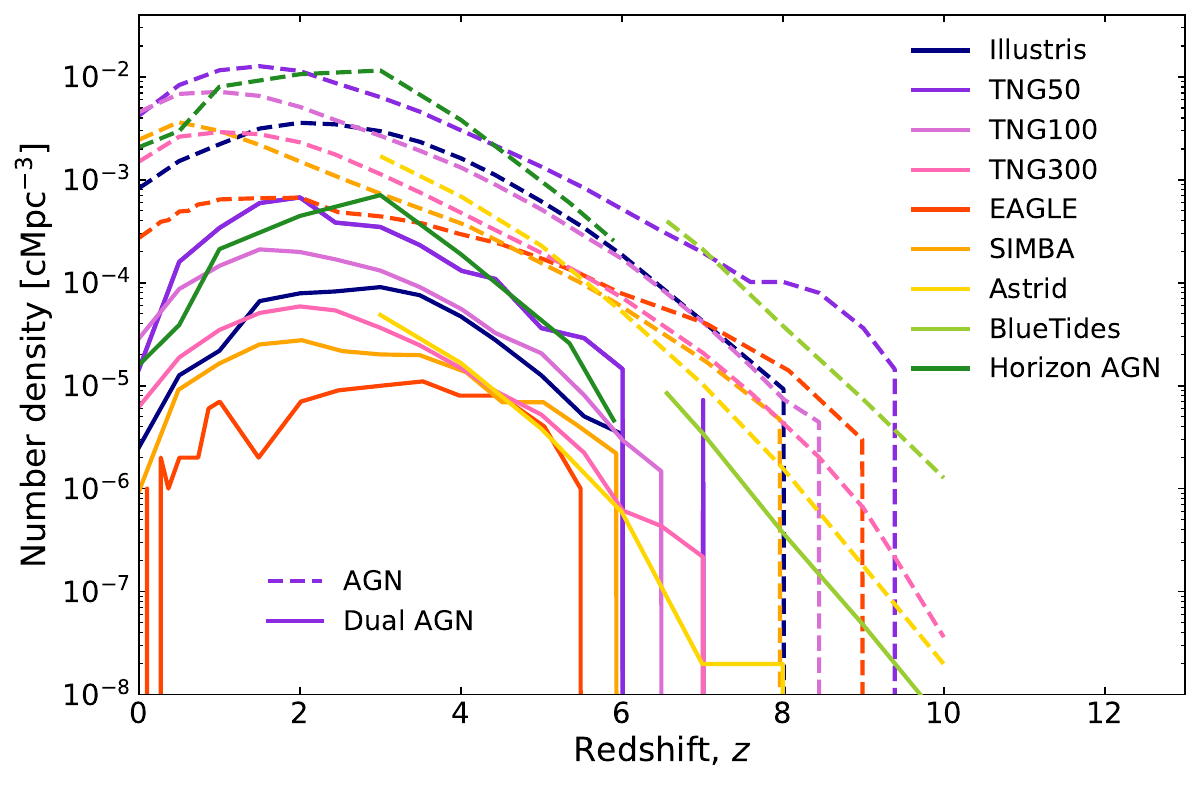}
    \caption{Number density of AGN (dashed lines) and DAGN (solid lines, identical to Fig.~\ref{fig:ndensity_fraction_z}) in simulations. Only AGN and DAGN with $L_{\rm bol}\geqslant 10^{43}\, \rm erg/s$ in galaxies with $M_{\star}\geqslant 10^{9}\, \rm M_{\odot}$ are considered.
    }
\label{fig:DAGN_follow_AGN}
\end{figure}

\begin{figure}
    \centering
    \includegraphics[scale=0.65]{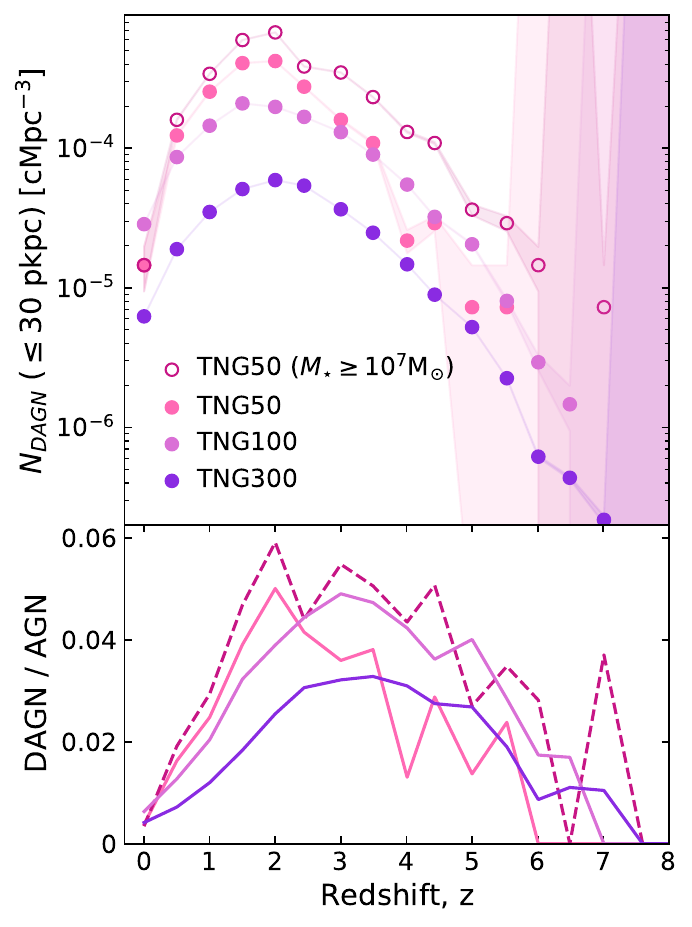}
    \caption{Same as Fig.~\ref{fig:ndensity_fraction_z} but only with the TNG simulations ($L_{\rm bol} \geqslant 10^{43}\, \rm erg/s$, $d \leqslant 30$ pkpc). 
    For the filled symbols or solid lines, only galaxies with $M_\star \geqslant 10^9 \, \rm M_\odot$ are considered (TNG50, TNG100, TNG300). Results for TNG50 when considering galaxies with $M_{\star} \geqslant 10^7 \, \rm M_\odot$ are added with purple open symbols and dashed lines. 
    }
    \label{fig:ndensity_resolution}
\end{figure}

\begin{figure*}
    \centering
    \includegraphics[scale=0.45]{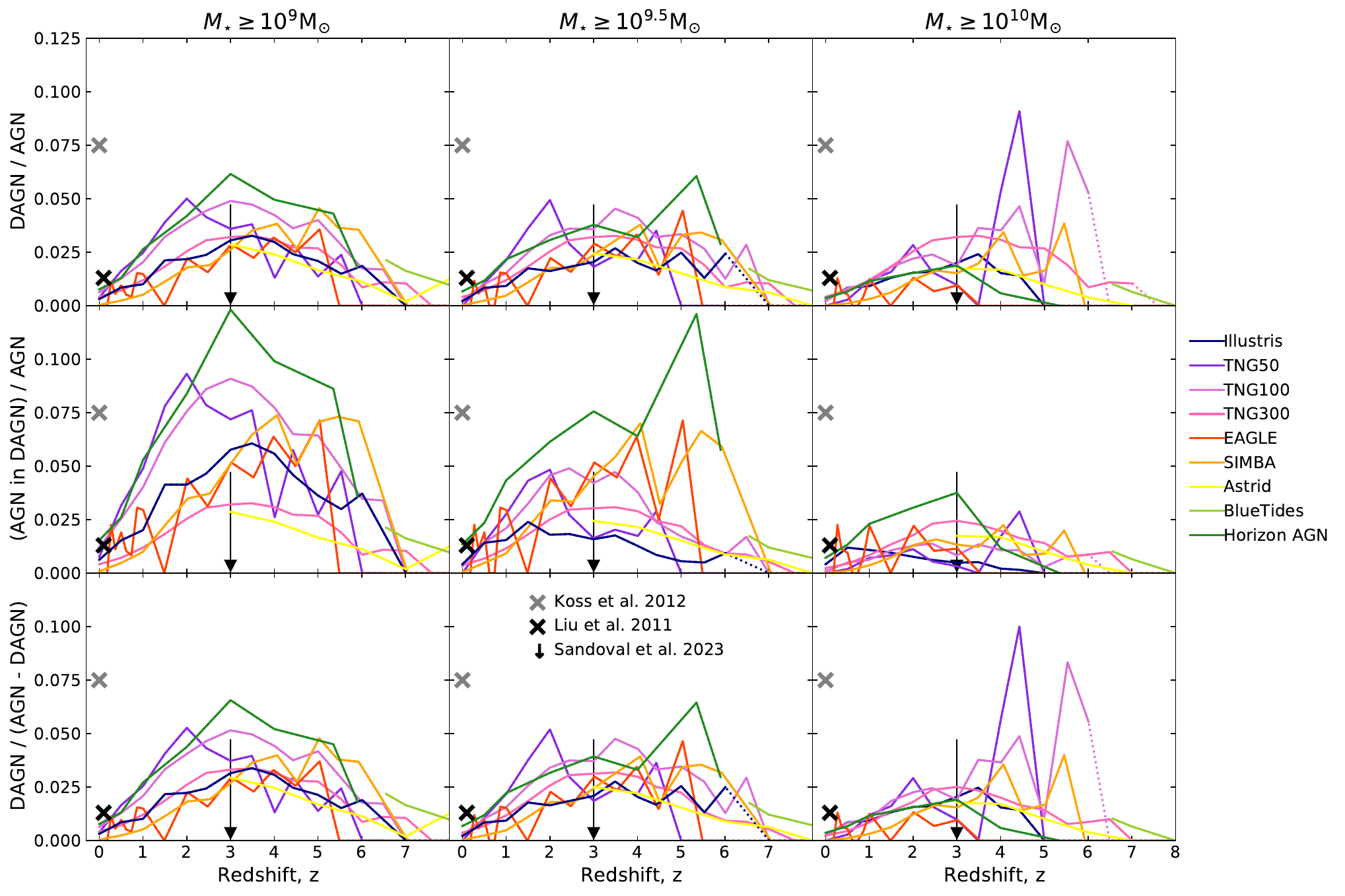}
    \caption{Comparison with the observational constraints from \citet{2011ApJ...737..101L, 2012ApJ...746L..22K} and upper limit from \citet{2023arXiv231202311S}; we report the constraints in all panels no matter the $M_{\star}$ cuts used for the simulations. {\it Top row:} Same definition of DAGN fraction as in the previous figures, i.e. the number of AGN pairs divided by total number of AGN in the simulations. {\it Middle row:} The fraction represents the number of AGN involved in a DAGN or multiple system divided by the total number of AGN in the simulations. {\it Bottom row:} The fraction is defined as the number of AGN pairs divided by the total number of AGN in the simulations to which we remove the number of identified pairs.}
\label{fig:comp_obs}
\end{figure*}

In Horizon-AGN, the larger number density of DAGN at $z=3$ is triggered by a large population of AGN at $z\sim 3$ \citep[see the jump in the normalization of the AGN luminosity function at $z\sim 3$ presented in Fig.~5 of][especially for $L_{\rm bol}\sim 10^{43}-10^{45}\, \rm erg/s$]{2021MNRAS.tmp.2905H}, and the overall larger number density of galaxies in that simulation. As discussed in our previous paper, this enhancement could also, to some extent, be driven by the creation of a new level of refinement in the hydrodynamical grid of Horizon-AGN at about this time in the simulation: gas distribution is better captured, and this can cause a sudden increase of MBH accretion rates. 
From Fig.~\ref{fig:DAGN_follow_AGN}, we note that in addition to the enhancement of AGN at $z\sim 3$ there is also an enhancement of DAGN, which means that with respect to the large number of AGN at that time, there is also proportionally more DAGN. This is shown by the larger fraction of DAGN at $z\sim 3$ in Fig.~\ref{fig:ndensity_fraction_z}. These enhancements in the fraction and number density of DAGN are only present when considering galaxies with $M_{\star}\geqslant 10^{9}\, \rm M_{\odot}$ and vanish for galaxies larger than $10^{9.5}\, \rm M_{\odot}$. This feature, when compared to observations, points towards too efficient MBH accretion in low-mass galaxies in Horizon-AGN.
In TNG50, including galaxies with $M_{\star}\geqslant 10^{7}\rm M_{\odot}$ also increases the number of DAGN (Fig.~\ref{fig:ndensity_resolution}) above the constraint of \citet{2023arXiv231202311S}. Characterising the properties of the DAGN host galaxies will thus become increasingly important to understand in which galaxies MBHs merge and the triggering of MBH fuelling in galaxies of different stellar masses. \\

\begin{figure}
    \centering
    \includegraphics[scale=0.62]{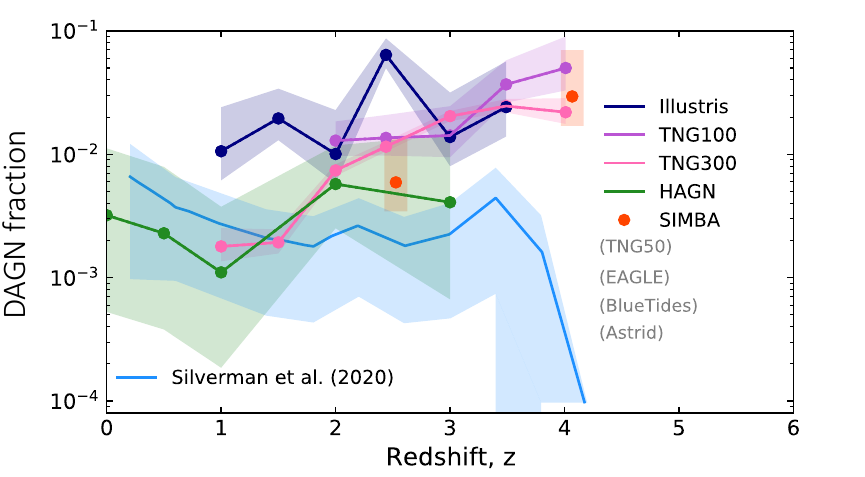}
    \caption{Comparison with the observational constraint from the Subaru/HSC program of \citet[][Subaru/HSC program, shaded region represents the 1$\sigma$ confidence interval]{2020ApJ...899..154S}. The fraction represents the number of AGN involved in dual or multiple-AGN systems divided by the total number of AGN in the simulations.
    We select DAGN with the same criteria: $d=5-30\, \rm pkpc$, primary AGN must have $L_{\rm bol}\geqslant 10^{45.3}\, \rm erg/s$, the secondary $L_{\rm bol}\geqslant 10^{44.3}\, \rm erg/s$, both powered by MBHs with $M_{\rm MBH}\geqslant 10^{8}\, \rm M_{\odot}$ in galaxies with $M_{\star}\geqslant 10^{10}\, \rm M_{\odot}$. We use 3-dimensional separation, while observations rely on projected separation, but it does not impact the results here. 
    No DAGN are found in 
    TNG50, EAGLE, and Astrid with these selection criteria and thus have null DAGN fractions. BlueTides was only run down to $z\geqslant 7$. 
    All the fractions shown are derived from at least 40 AGN at each redshift (in the range $z=0-4$), while the number of DAGN (if any) varies from 1 to 6. For the simulations, error bars represent Poisson errors for low counts (exact Poisson confidence intervals using chi-square distribution).}
\label{fig:silverman2020}
\end{figure}

\begin{figure*}
    \centering
    \includegraphics[scale=0.58]{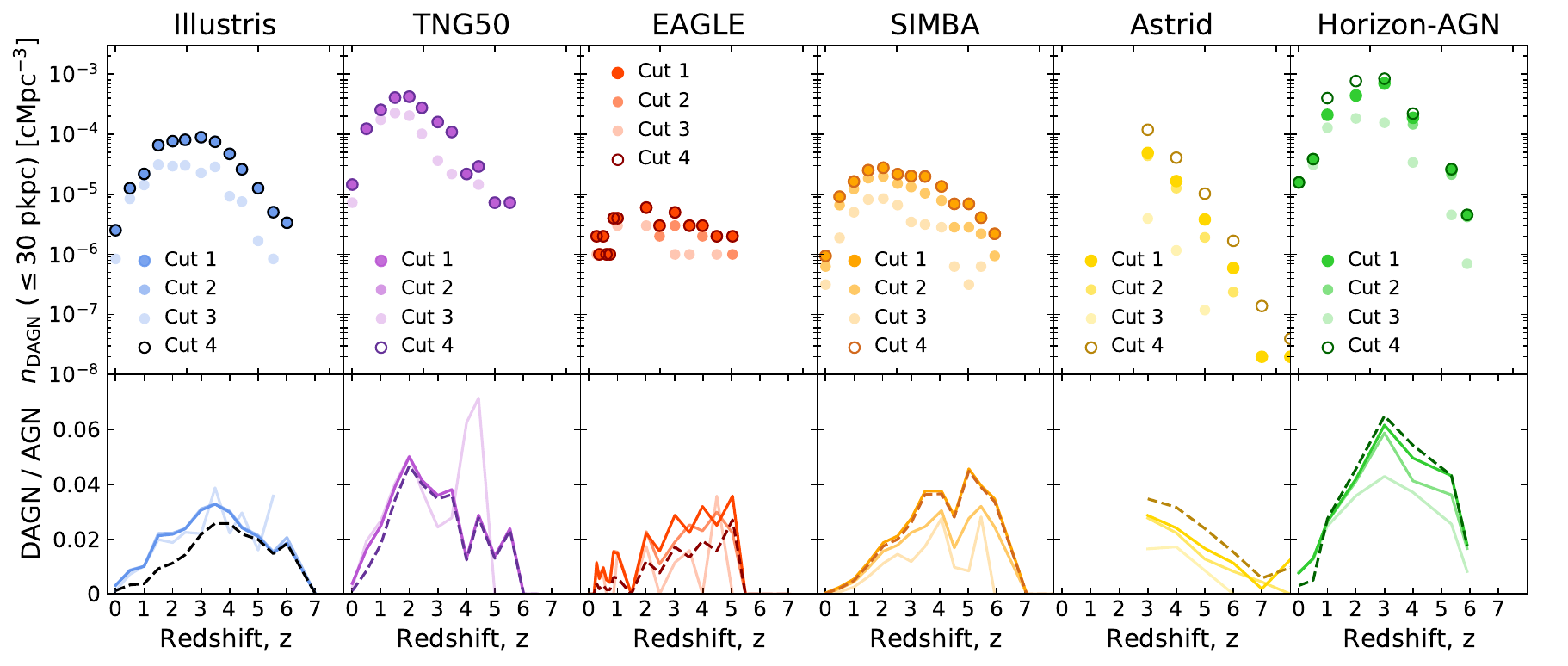}
    \caption{Same as Fig.~\ref{fig:ndensity_fraction_z} with different selection criteria to identify the DAGN produced in the simulations. The results shown with the stronger colour (``Cut 1'') are identical to those in Fig.~\ref{fig:ndensity_fraction_z}: M$_\star \geqslant 10^9$ M$_\odot$, L$_{\rm bol}\geqslant 10^{43}$ erg/s.
    With ``Cut 2''/ ``Cut 3'' (lighter colours), we consider only massive MBHs with M$_{\rm MBH}\geqslant 10^6\, /\, 10^{7}$ M$_\odot$ (and again M$_\star \geqslant 10^9$ M$_\odot$, L$_{\rm bol} \geqslant 10^{43}$ erg/s).
    With ``Cut 4'' (dashed lines, open symbols), we consider fainter DAGN with L$_{\rm bol} \geqslant 10^{42}$ erg/s (M$_\star \geqslant 10^9$ M$_\odot$, all MBH masses).
    }
\label{fig:DAGN_fraction_variation2}
\end{figure*}

We pursue our comparison with high-redshift observations with the work of \citet{2020ApJ...899..154S}. The authors derive a fraction of luminous DAGN for $z\leqslant 3.5$ with a separation of $5-30 \, \rm kpc$. DAGN are identified as one AGN with $L_{\rm bol}\geqslant 10^{45}\, \rm erg/s$ and the second with at least $10\%$ of the former AGN, for MBHs of $M_{\rm MBH}\geqslant 10^{8}\, \rm M_{\odot}$ and $M_{\star}\geqslant 10^{10}\, \rm M_{\odot}$. The authors derive a fraction of  $0.26 \pm 0.18\%$ with no evolution with redshift.
Likely due to the different quasar selection, \citet{2020ApJ...899..154S} finds a lower DAGN fraction (by a factor of 4) at $z < 1$ than \citet{2011ApJ...737..101L} and \citet{2012ApJ...746L..22K}. A similar fraction of $\sim 6.2\pm 0.5\times 10^{-4}$ is reported in \citet{2023ApJ...943...38S} for brighter and unobscured systems with $L_{\rm bol}>10^{45.8}\, \rm erg/s$ at $1.5<z<3.5$.
In Fig.~\ref{fig:silverman2020}, we report the fractions of luminous DAGN from the simulations with the selection criteria presented in \citet{2020ApJ...899..154S}, i.e., a primary AGN with $L_{\rm bol}\geqslant 10^{45.3}\, \rm erg/s$, a secondary 10 times fainter ($L_{\rm bol}\geqslant 10^{44.3}\, \rm erg/s$), both powered by MBHs with $M_{\rm MBH}\geqslant 10^{8}\, \rm M_{\odot}$ in galaxies with $M_{\star}\geqslant 10^{10}\, \rm M_{\odot}$.
To match the definition used in \citet{2020ApJ...899..154S}, our fractions in Fig.~\ref{fig:silverman2020} represent the number of AGN involved in DAGN (or multiple systems) divided by the total number of AGN in the simulations.

We find that several of the simulations (TNG50, EAGLE, Astrid) do not produce any DAGN with the above criteria, the limiting factor being the high luminosity of the primary AGN in \citet{2020ApJ...899..154S} \citep[or both AGN in][]{2023ApJ...943...38S} or the small simulated volume for TNG50. 
We find a relatively good agreement with the observational constraints for the simulations that do produce luminous DAGN (Illustris, TNG100/TNG300, SIMBA, Horizon-AGN), 
particularly at $z\leqslant3$. 
However, some simulations do predict an evolution with redshift (e.g., the fractions of the TNGs drop by about 0.5-1 dex) that is not seen in the observations (or with much lower amplitude). Our results are limited by small number statistics as the number of DAGN at each redshift shown in Fig.~\ref{fig:silverman2020} varies from 1 to 6 DAGN, our error bars represent Poisson errors for low counts (i.e., exact Poisson confidence intervals using chi-square distribution).
As already discussed in \citet{2021A&A...654A..79G} with TNG300, we find that the simulations do not produce environments dense enough to nourish bright and massive DAGN at high redshifts (i.e., beyond $z=4$ in Fig.~\ref{fig:silverman2020}). Interestingly, a similar drop in the DAGN fraction in the observations suggests that DAGN involving two bright AGN (including a quasar) are very rare or do not exist even in the dense environments of the Universe at high redshifts.   
We caution here that the exact assessment of the agreement between simulations and observational constraints is, of course, subject to uncertainties (e.g., simulations do not perfectly match observational constraints on the AGN luminosity function) and the DAGN fractions is sensitive to the exact selection criteria, as demonstrated and described in \citet{2022MNRAS.514..640V} for the same constraint \citet{2020ApJ...899..154S}.

A contrasting result was recently presented in \citet{2023arXiv231003067P}. The authors identified a triple AGN and 3 dual AGN (plus an additional candidate) at $z\sim 3.5$ on a sample of 17 AGN ($2<z<6$). These discoveries were made possible with the JWST NIRSpec and MUSE, or NIRSpec/IFS.
The AGN of the multiple systems are within $L_{\rm bol}\sim 10^{44}-10^{46}\, \rm erg/s$, separated by projected distances $\rm < 30 \, kpc$ and hosted by galaxies with $M_{\star}\sim 10^{8}-10^{11}\, \rm M_{\odot}$ (and probably central galaxies, not satellite). The corresponding fraction of $20-30\%$ multiple AGN systems at $z\sim 3$ is interesting since much higher than the $\leqslant 6\%$ found in the simulations studied in this paper (Fig.~\ref{fig:ndensity_fraction_z_44} for $L_{\rm bol}\geqslant 10^{44}\, \rm erg/s$). Future detections with new facilities such as JWST will shed new light on the fraction of multiple AGN and their number density in volume-limited surveys.

Finally, the detections of dual quasars (and candidates) have been extended to $z\geqslant5$ \citep{2023AJ....165..191Y,2021ApJ...921L..27Y}, with separation of kpc scale. Computing the fraction of dual systems composed of massive MBHs at such redshifts still suffers from too low statistics in both observations and simulations.

\subsection{Fraction of dual AGN as a function of $M_{\rm MBH}$ and $L_{\rm bol}$ and comparison with previous studies}
\label{sec:study_dependencies}
To facilitate comparisons with previous works and observational results, we derive the fraction of DAGN for different cuts in $M_{\rm MBH}$ and $L_{\rm bol}$ in Fig.~\ref{fig:DAGN_fraction_variation2}.
Overall and for most of the simulations, considering MBHs with any MBH masses or $M_{\rm MBH}\geqslant 10^{6}$ or $10^{7}\, \rm M_{\odot}$ 
impacts the number densities and fractions of DAGN
across cosmic times (see the different coloured lines and symbols in Fig.~\ref{fig:DAGN_fraction_variation2}) while mostly conserving their trends with redshift. Variations in number densities are within a dex in Illustris, TNG50, EAGLE, Horizon-AGN, and SIMBA. They can be slightly larger at some redshifts for Astrid (when considering the most restricting limit of $M_{\rm MBH}\geqslant 10^{7}\, \rm M_{\odot}$).

\begin{figure*}
    \centering
    \includegraphics[scale=0.7]{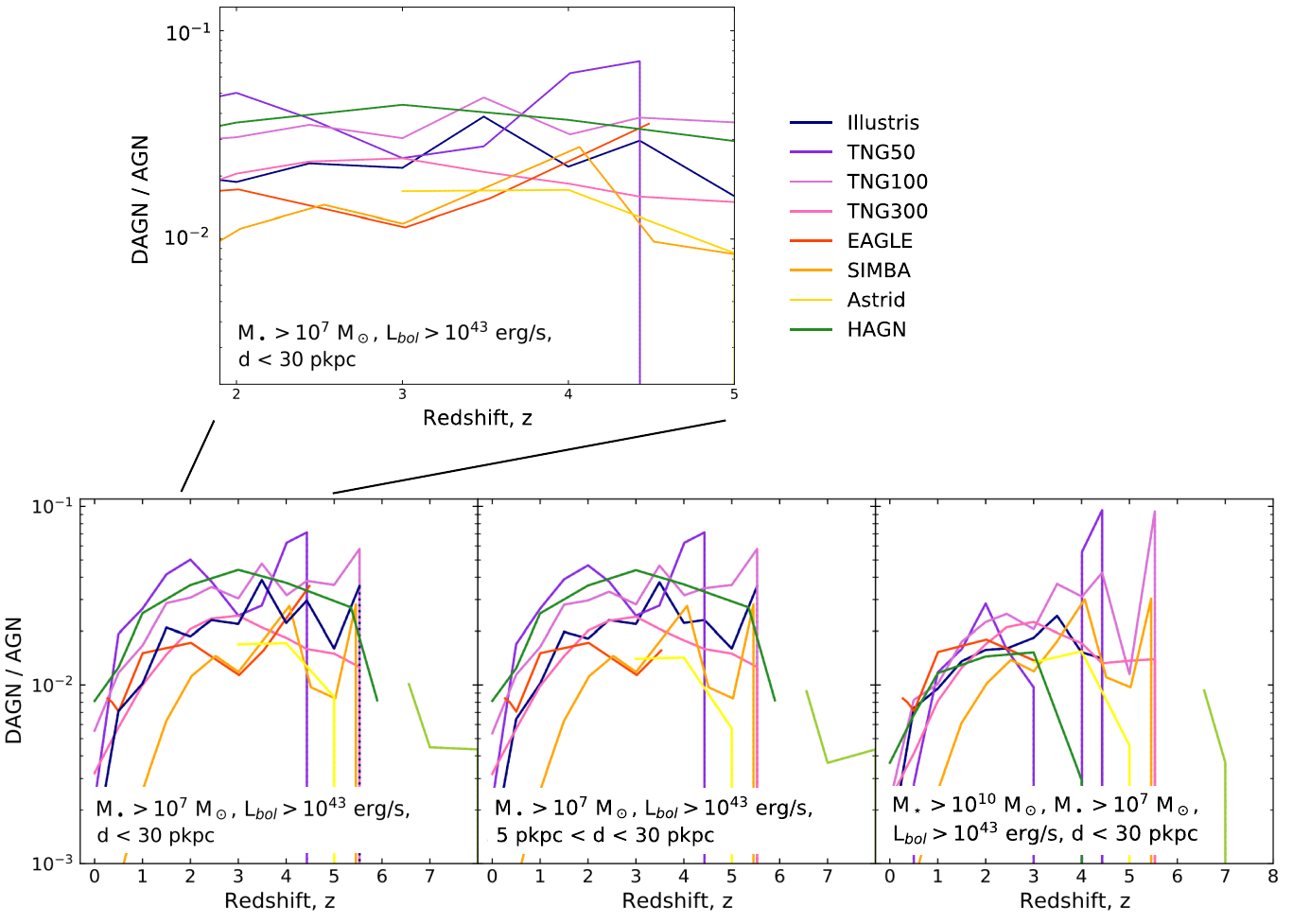}
    \caption{{\it Top panel}: Fractions of DAGN produced in cosmological simulations for MBHs with $M_{\rm MBH}\geqslant 10^{7}\, \rm M_{\odot}$ in galaxies of $M_{\star}\geqslant 10^9\, \rm M_{\odot}$.
    {\it Bottom panels}: DAGN fractions for the full redshift range (left panel), for $d=5-30 \, \rm kpc$ (middle panel, almost identical to the left panel), and for massive galaxies with $M_{\star}\geqslant 10^{10}\, \rm M_{\odot}$(right panel).}
    \label{fig:DAGN_fraction_variation_2}
\end{figure*}

The impact of the MBH mass cut is the result of a complex combination of the MBH mass function \citep[see Fig.~13,\, 15 in][]{2021MNRAS.503.1940H}, ability of the MBHs to accrete gas and trigger AGN with $L_{\rm bol}\geqslant 10^{43}\, \rm M_{\odot}$, and the probability of having close AGN.
For the former, both the amplitude of the mass function (i.e., the number of MBHs reaching a mass of, e.g. $10^{7}\, \rm M_{\odot}$) and the shape of the function are important. Some simulations, such as SIMBA (at all redshifts) or Illustris (at low redshifts)
have a relatively flat mass function for the mass range $10^{6}-10^{7}\, \rm M_{\odot}$: having a mass cut for DAGN at $10^{6}$ or $10^{7}\, \rm M_{\odot}$ can thus quite strongly change the number density of DAGN. Other simulations with a mass function peaking at $10^{7}\, \rm M_{\odot}$ are in theory less impacted by a mass cut of $10^{6}$ or $10^{7}\, \rm M_{\odot}$ (since the highest number of MBHs peak at $10^{7}\, \rm M_{\odot}$). For most simulations in Fig.~\ref{fig:DAGN_fraction_variation_2}, number densities for different MBH mass cuts are most affected at high redshifts, which reflects the rarity of massive MBHs with, e.g. $10^{7}\, \rm M_{\odot}$ 
(i.e. the lower amplitude of the MBH mass function at high redshift).

The last selection criterion we tested in Fig.~\ref{fig:DAGN_fraction_variation2} is the luminosity of the DAGN. Open symbols and dashed lines show the fractions for fainter AGN ($L_{\rm bol}\geqslant 10^{42}\, \rm erg/s$) than before considered ($L_{\rm bol}\geqslant 10^{43}\, \rm erg/s$)\footnote{Number densities and fractions for brighter DAGN with $L_{\rm bol}\geqslant 10^{44}\, \rm erg/s$ are displayed in Fig.~\ref{fig:ndensity_fraction_z_44}.}. As expected, considering DAGN triggered by fainter AGN leads to identical or larger DAGN number densities.
However, the effect on DAGN fractions is not trivial. It can go either way, depending on the impact on the AGN population produced in the simulations. In Illustris, for example, the lower luminosity limit encompasses more AGN than DAGN, which results in a lower fraction of DAGN. We identify the same trend in TNG, EAGLE. There is no variation for SIMBA, and we find an increase of the fraction in Astrid and Horizon-AGN \citep[see also Fig.~17 in][]{2022MNRAS.514..640V}.

An interesting feature presented in \citet[][their Fig.~2, Astrid simulation]{2023MNRAS.522.1895C} is the constant fraction of AGN with $M_{\rm MBH}\geqslant 10^{7}\, \rm M_{\odot}$ involved in DAGN in the redshift range $z=2-5$ (and even more so for DAGN separations of $d\geqslant 5\, \rm pkpc$). We derive the DAGN fractions produced by all the simulations with such criteria and display the results in Fig.~\ref{fig:DAGN_fraction_variation_2}. 
For that figure, we consider all DAGN separated by $d\leqslant 30\, \rm kpc$ found in distinct galaxies; enforcing $d=5-30\, \rm pkpc$ (shown in the bottom of Fig.~\ref{fig:DAGN_fraction_variation_2} along with another version considering DAGN in more massive galaxies only) with our above criteria does not impact the fractions. Although the DAGN fraction is indeed close to being constant in Astrid and other simulations for massive MBHs \citep[TNG100/TNG300, Illustris, Horizon-AGN, see also][]{2022MNRAS.514..640V}, some simulations such as TNG50 present stronger variations in the redshift range $z=2-5$. 
Our analysis also highlights three important aspects. First, considering a broader redshift range is paramount to understanding the full cosmic evolution of DAGN and identifying the redshift at which AGN most cluster in multiple systems. Second, observing only massive MBHs with $M_{\rm MBH}\geqslant 10^{7}\, \rm M_{\odot}$ may not reflect the DAGN fraction for the entire MBH population, according to several simulations. For example, the peak in the DAGN fraction at $z\sim 4$ produced in TNG50 for $M_{\rm MBH}\geqslant 10^{7}\, \rm M_{\odot}$ does not mean that there is also an enhancement of DAGN at the same redshift when considering all DAGN powered by $M_{\rm MBH}\geqslant 10^{6}\, \rm M_{\odot}$ (Fig.~\ref{fig:DAGN_fraction_variation2}, Fig.~\ref{fig:DAGN_fraction_variation_2}).
Third, the hosts in which DAGN are found have a strong impact on the DAGN fraction (right panel in Fig.~\ref{fig:DAGN_fraction_variation_2}) and can make difficult any comparisons with observations if the galaxy stellar masses are unconstrained. So far, the characterisation of the AGN population across cosmic times has been mostly based on observations in galaxies of $M_{\star}\geqslant 10^{10}\, \rm M_{\odot}$ and likely triggered by MBHs of at least $10^{7}\, \rm M_{\odot}$. With these criteria, we find strong variations in the predicted DAGN fractions. Results from observations with these criteria should not be extrapolated to populations of DAGN triggered by lower-mass MBHs and/or less massive galaxies.

\begin{figure*}
    \centering
    \includegraphics[scale=0.6]{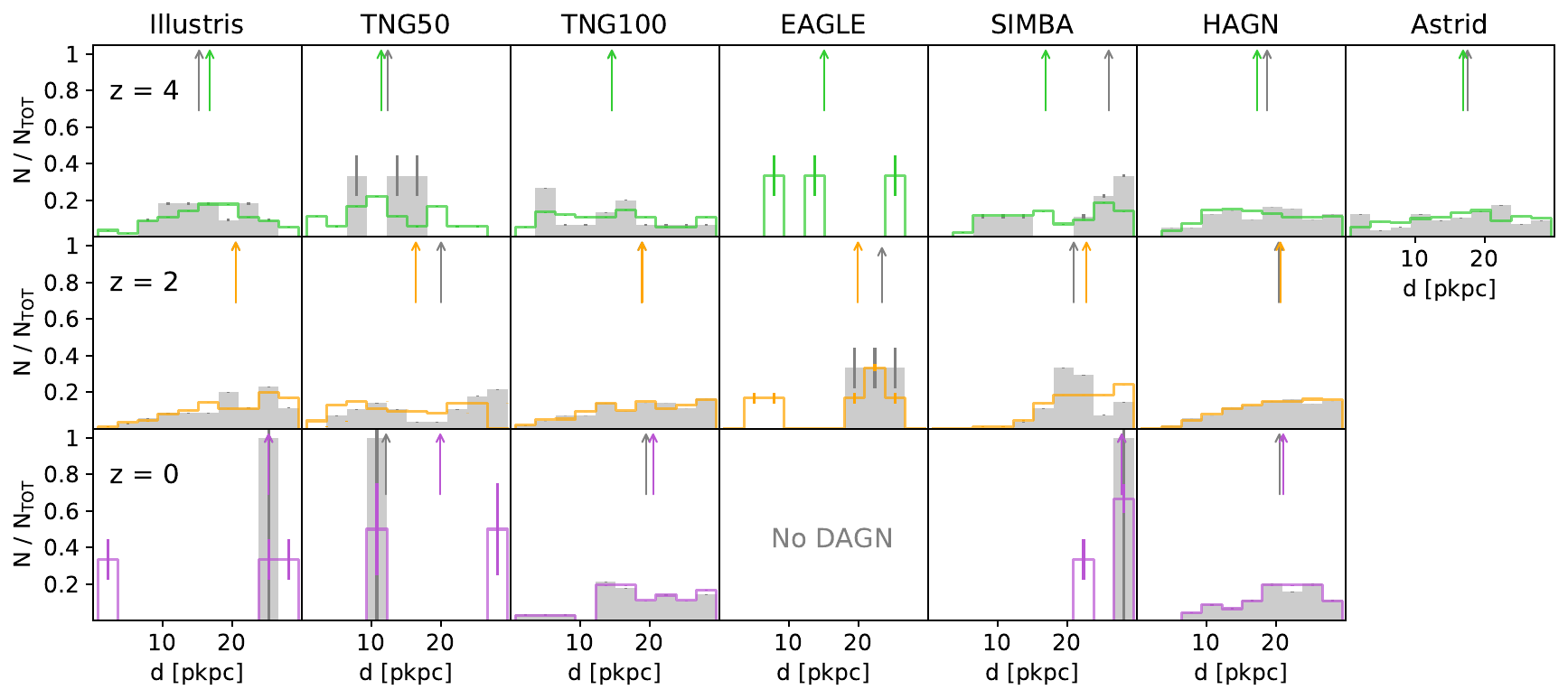}
    \caption{Normalized distribution of the 3-dimensional separations of DAGN (coloured histograms) and DAGN with M$_{\rm MBH} \geqslant 10^7$ M$_\odot$ (grey filled distributions)  
    for redshift $z = 4,\, 2,\,0$. Poisson error bars have been included on top of the bar plots. Each distribution was normalised by the total number of pairs identified for each study. Only galaxies with $M_\star \geqslant 10^{9} \rm M_\odot$ are considered, and AGN with $L_{\rm bol}\geqslant 10^{43}\, \rm erg/s$. Black arrows indicate the median value of each distribution.} 
    \label{fig:hist_dual_d_60}
\end{figure*}

\subsection{Number densities of multiple AGN}
\label{sec:multiple_AGN}
This section investigates the occurrence of triple (or more) AGN in cosmological simulations. We keep the same criteria to identify multiple pairs (any MBH masses, $L_{\rm bol}\geqslant10^{43}\, \rm erg/s$, $M_{\star}\geqslant 10^{9}\, \rm M_{\odot}$, $d\leqslant 30\,\rm pkpc$).
Generally speaking, triple systems are more than an order of magnitude less represented in simulations than DAGN. First, at high redshift in BlueTides, the number density at $z=6.5$ drops from $10^{-5}\, \rm cMpc^{-3}$ for DAGN to $5\times 10^{-7}\, \rm cMpc^{-3}$ for triple AGN, and to $5\times 10^{-8}\, \rm cMpc^{-3}$ for systems that involve more than 3 AGN. At $z=3$, the density evolves as $7\times 10^{-4}$, $10^{-4}$, $3\times 10^{-5}\, \rm cMpc^{-3}$ in Horizon-AGN for DAGN, triple AGN, and more \citep[see also][]{2022MNRAS.514..640V}. 
This is the simulation with the highest number of AGN and multiple systems, with a relatively high number of systems with more than 3 AGN. In Astrid, we find an order of magnitude less triplets ($4\times 10^{-6}\, \rm cMpc^{-3}$), and two orders of magnitude less systems with more than 3 AGN ($3\times 10^{-7}\, \rm cMpc^{-3}$) with respect to DAGN at $z=3$ ($7\times 10^{-5}\, \rm cMpc^{-3}$). Although Astrid produces DAGN at $z\geqslant 5$, all of its multiple systems are found at later times.
The redshift evolution of the number densities for triplets and more AGN is conserved in all the simulations studied here. In particular, these systems peak at the same time as DAGN. As we concluded for DAGN, multiple AGN also become rarer towards $z=0$, and some simulations predict none.

In terms of number density, our results agree with the dedicated study of \citet{2020MNRAS.492.5620B}
for the large-scale MassiveBlackII simulation in the redshift range $0\leqslant z \leqslant4$.
The authors find a number density of triple and quadruple AGN systems with $L_{\rm bol}>10^{44}\, \rm erg/s$ ranging in $10^{-5}-10^{-6}\, \rm h^{3}Mpc^{-3}$. They found the highest number of such systems in the redshift range $1.5\leqslant z\leqslant 2.5$, which is in good agreement with our findings with our luminosity cut of $L_{\rm bol}\geqslant10^{43}\, \rm erg/s$. Using a more restrictive cut of $\geqslant 10^{44}\, \rm erg/s$ as presented in Fig.~\ref{fig:ndensity_fraction_z_44} further reduces the number of multiple systems.

\subsection{Separation of simulated dual AGN}
\label{sec:separation}
We present in Fig.~\ref{fig:hist_dual_d_60} the normalized distribution of DAGN as a function of 3-dimensional separation (solid coloured lines) 
for redshifts $z = 4,\,2,\, 0$. 
As before, only galaxies with $M_\star \geqslant 10^{9}\,\rm M_\odot$ are included ($M_\star \geqslant 10^{7} \, \rm M_\odot$ for TNG50) and MBHs are considered as AGN if $L_{\rm bol} \geqslant 10^{43}$ erg/s. In each panel, an arrow indicates the median of the distribution.

Although with significant differences, all simulations produce relatively flat distributions of DAGN separation with a smaller number of pairs at small separations ($d < 10$ pkpc) for any redshift. This is because we only consider in this paper DAGN in distinct galaxies; single-galaxy DAGN are generally separated by $d < 10$ pkpc \citep[as detailed in][]{2022MNRAS.514..640V,2023MNRAS.522.1895C}. In addition, in most cosmological simulations MBHs would be {\it numerically} merged if separated by only a few resolution elements, for example, a couple of kpc; no DAGN should then be identified at $\rm \sim kpc$ separation in these simulations.   
Most of the DAGN candidates identified so far beyond the local Universe up to $z=2.5$ are separated by $d\geqslant 10\, \rm pkpc$. This limit of $d \sim 10$ pkpc corresponds to our current observational capacity at those intermediate redshifts. As shown in Fig.~\ref{fig:ndensity_fraction_z} (described in more detail in Section~\ref{sec:detectability}), most candidates below that limit were identified with the high-sensitivity HST. The flat distributions that we find here for $d \geqslant 10$ pkpc are in good agreement with \citet[][their Fig.~3]{2023MNRAS.522.1895C} at $z=2-3$, although the authors only consider DAGN powered by $M_{\rm MBH}\geqslant 10^{7}\, \rm M_{\odot}$ (which correspond to our grey filled histograms in Fig.~\ref{fig:hist_dual_d_60}).

We find that the median separation increases with time for all the simulations. At high redshift (e.g., $z=4$ shown in the top panels of Fig.~\ref{fig:hist_dual_d_60}), DAGN are on average separated with a median of $\sim 15\, \rm pkpc$. By $z=2-0$, DAGN are more separated with a median in the range $\geqslant 20\, \rm pkpc$. This is in agreement with the results presented in \citet[][their Fig.~3]{2019MNRAS.483.2712R} for EAGLE, although with a slightly different DAGN selection.
The low statistics that we find for $z=0$ (lower predicted DAGN number than the current number of observed DAGN candidates, see Fig.~\ref{fig:ndensity_fraction_z}) prevent us from drawing strong conclusions on the typical separation of DAGN in the local Universe.

We also show the distributions for DAGN powered by more massive MBHs with $M_{\rm MBH}\geqslant 10^{7}\, \rm M_{\odot}$ (grey filled histograms) that are more generally found in observations and find consistent results (although with lower statistics).

\begin{figure*}
    \centering
    \includegraphics[scale=0.6]{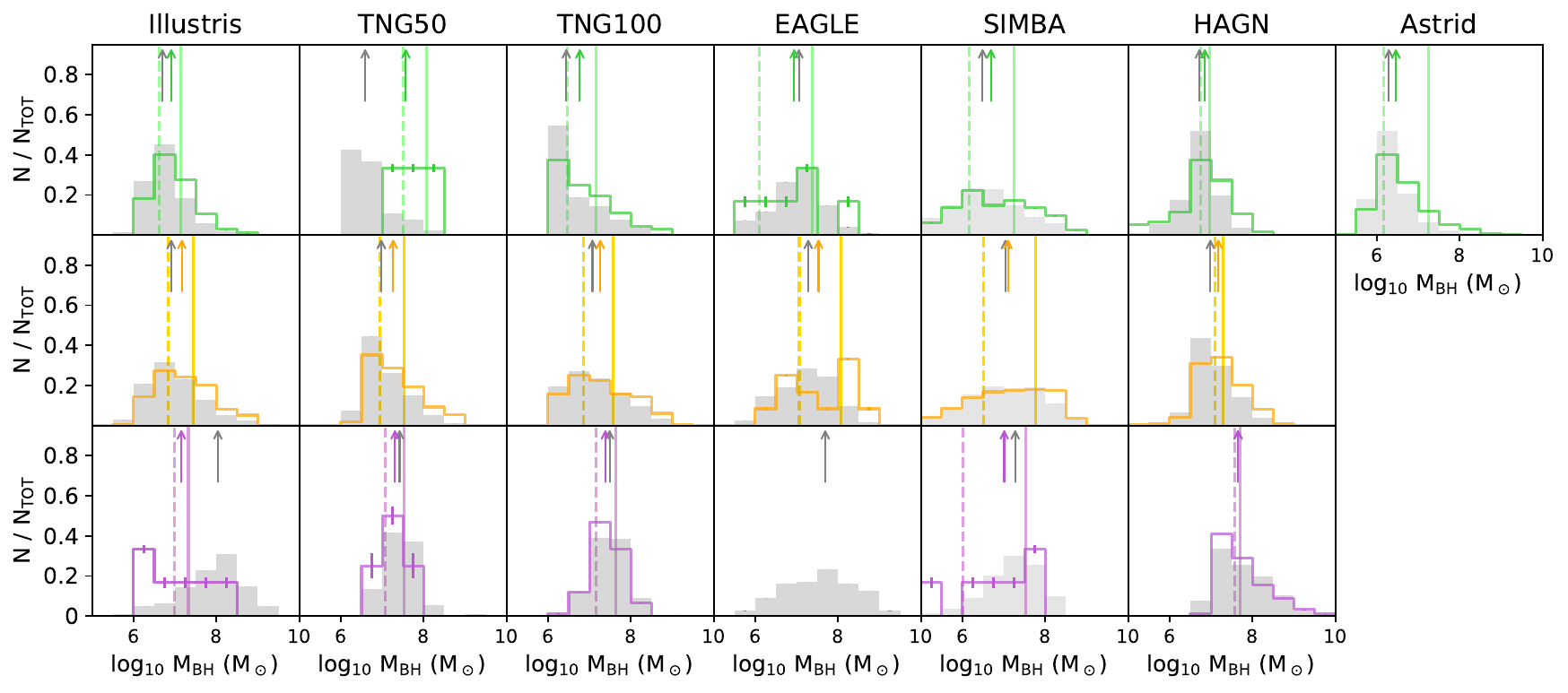}
    \caption{Normalized distribution of DAGN (coloured histograms) and AGN (grey distributions) as a function of their MBH mass for $z= 4,\, 2,\,0$. 
    Only galaxies with $M_\star \geqslant 10^{9}\,\rm M_\odot$ are considered, and AGN with $L_{\rm bol}\geqslant 10^{43}\, \rm erg/s$. 
    Poisson error bars are included on top of the bar plots. Each distribution has been normalised according to the total number of pairs identified in each simulation. 
    Arrows indicate the median value of each distribution.
    The medians of the primary (solid lines) and secondary (dashed lines) MBHs of the DAGN are shown as vertical coloured lines. 
    }
    \label{fig:hist_Mbh}
\end{figure*}

\begin{figure*}
    \centering
    \includegraphics[scale=0.6]{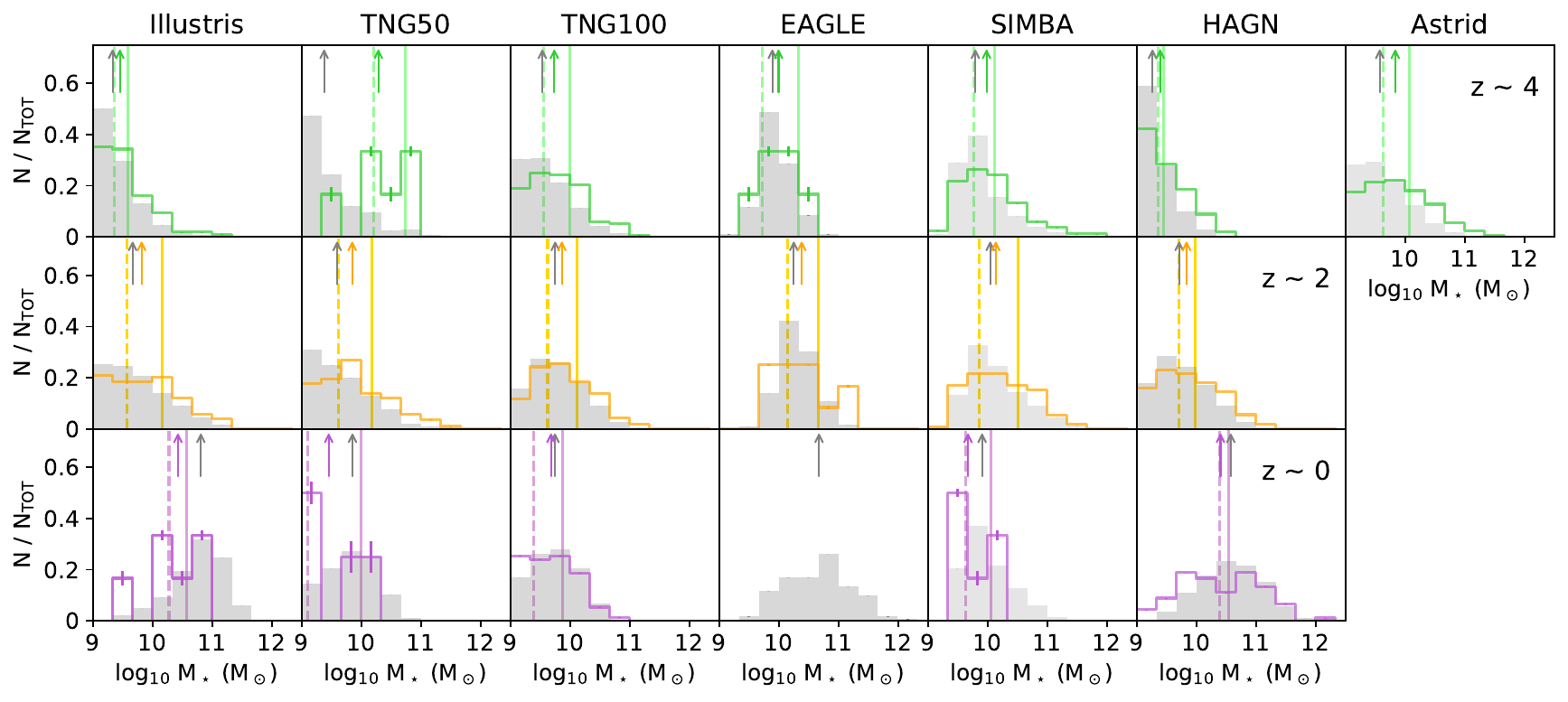}
    \caption{Normalized distribution of DAGN (coloured histograms) and AGN (grey distributions) as a function of their host galaxy stellar mass for $z= 4,\, 2,\,0$.  Same $M_{\star}$, $L_{\rm bol}$ selection as previous figures. Poisson error bars are included on top of the bar plots. Each distribution has been normalised according to the total number of pairs identified in each simulation. 
    Arrows indicate the median value of each distribution.
    The medians of the primary (solid lines) and secondary (dashed lines) MBHs of the DAGN are shown as vertical coloured lines. 
    DAGN are hosted in more massive galaxies than AGN at high redshift, while there is no consensus at $z=0$ in the simulations. We find that the primary MBHs are hosted by the most massive galaxies of the pairs in all the simulations and at all redshifts. 
    }
    \label{fig:hist_Mstar}
\end{figure*}

\begin{figure*}
    \centering
    \includegraphics[scale=0.6]{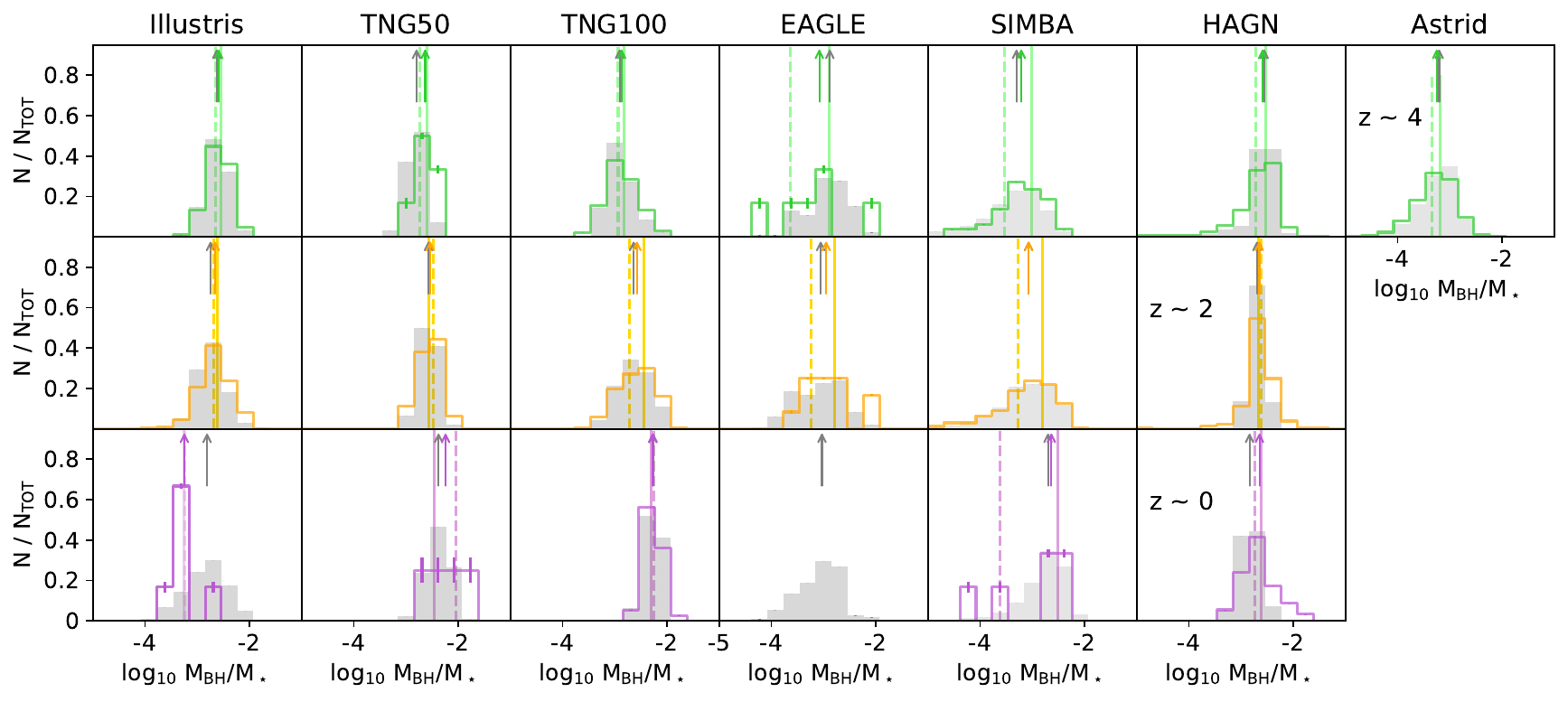}
    \caption{Normalized distributions of DAGN (coloured histograms) and AGN (grey filled distributions) 
    as a function of $M_{\rm MBH}/M_{\star}$ ratios at $z= 0$. Same $M_{\star}$, $L_{\rm bol}$ selection as previous figures. Arrows indicate the median value of each distribution. 
    Poisson error bars are included on top of the bar plots. Each distribution has been normalised according to the total number of pairs identified in each simulation. 
    No significant differences are identified at higher redshift ($z\geqslant 2$). The medians of the primary (solid lines) and secondary (dashed lines) MBHs of the DAGN are shown as vertical coloured lines; while primary MBHs seem to be slightly overmassive with respect to their galaxies (and secondary undermassive) at high redshift, the difference is small. 
    }
    \label{fig:hist_Mbh_Mstar}
\end{figure*}

\begin{figure*}
    \centering
    \includegraphics[scale=0.6]{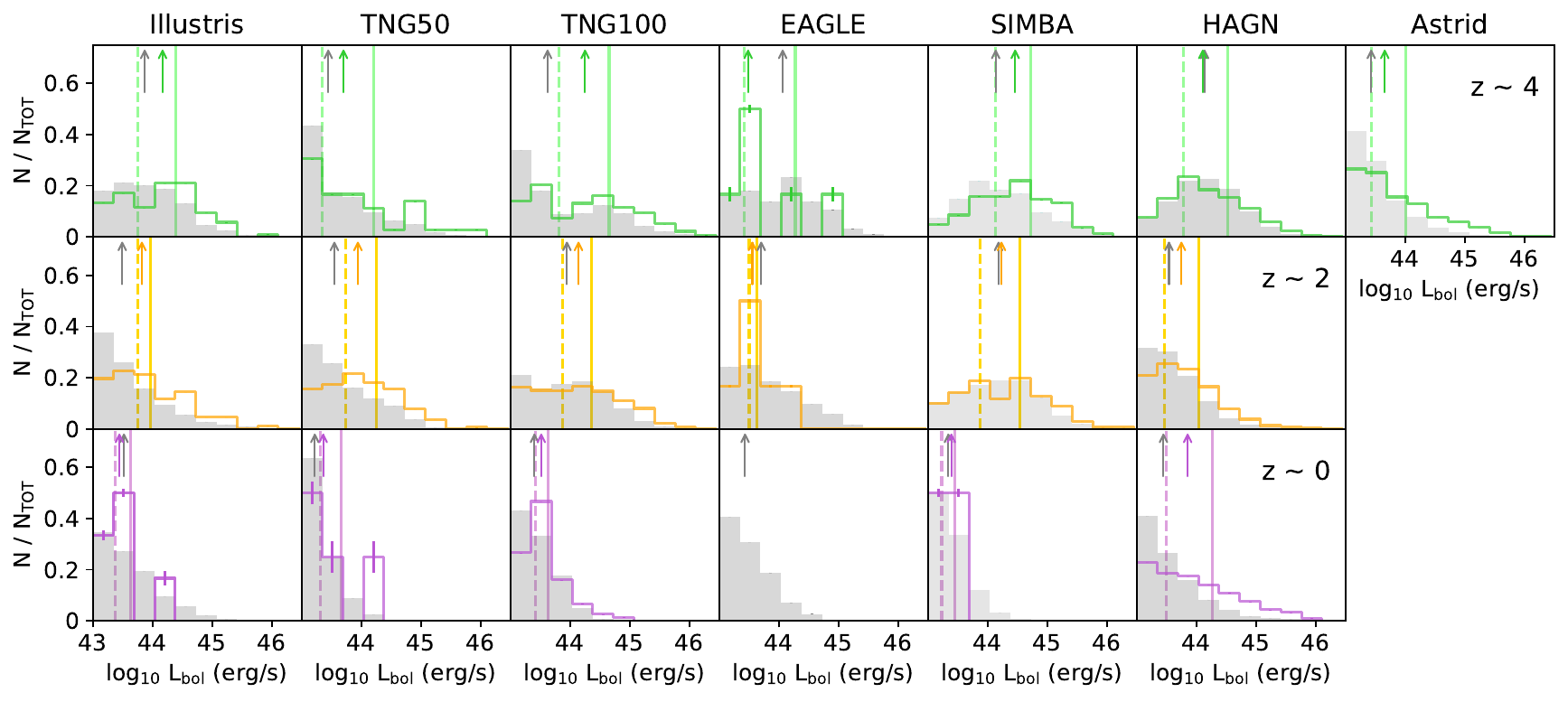}
    \caption{Distribution of DAGN (coloured histograms) and AGN (grey filled distributions) bolometric luminosities for $z= 4,\, 2,\,0$, normalized to the total number of objects in the histograms. Same $M_{\star}$, $L_{\rm bol}$ selection as previous figures. Arrows indicate the median of the distributions. 
    The medians of the primary (solid lines) and secondary (dashed lines) MBHs of the DAGN are shown as vertical-coloured lines. In all the simulations, except for EAGLE and Horizon-AGN (at high redshift), DAGN are, on average, brighter than the full AGN population. Primary MBHs are, on average, the brightest AGN of the pairs.}
    \label{fig:hist_Lbol}
\end{figure*}

\section{Which galaxies/MBH/environment power dual AGN?}
\label{sec:galaxies}

In this section, we explore the properties of the MBHs that pair and those of their host galaxies and how they vary with time and with the subgrid modelling of each cosmological simulation. Most of the figures discussed in the following include histograms of DAGN when taking together primary (most massive) and secondary MBHs (less massive of a given pair) and show an indication of the median of the distribution for the primary MBHs as well as the secondary MBHs.

\subsection{MBH and host galaxy properties of dual AGN}
\subsubsection{MBH and host stellar masses}
Although with differences, in most simulations DAGN (taking primary and secondary MBHs together) are on average powered by MBHs with $\sim 10^{6.5}-10^{7}\, \rm M_{\odot}$ (shown in Fig.~\ref{fig:hist_Mbh}). 
Their median $M_{\rm MBH}$ moves towards more massive MBHs at later times for all simulations. 
This is a natural consequence of AGN evolution across cosmic times; at later times, the presence of more massive MBHs can trigger more AGN with $L_{\rm bol}\geqslant 10^{43}\, \rm erg/s$ in the Bondi accretion formalism (under the assumption that gas is available). 
Interestingly, the median of the DAGN moves towards more massive MBHs with time but is less than the full AGN population. So, at high redshift, DAGN are always powered by slightly more massive MBHs than the AGN population, but at $z=0$ the differences vanish or DAGN are powered by less massive MBHs than the AGN. The greatest effect is identified in Illustris (DAGN have a median of $10^{7}\, \rm M_{\odot}$, and AGN $10^{8}\, \rm M_{\odot}$) and SIMBA at $z=0$.

Going further, two simulations are particularly interesting at $z=4$. First, while the population of AGN in EAGLE extends from $10^{6}$ to $10^{8}\, \rm M_{\odot}$ at that redshift, most DAGN are powered by MBHs of $\geqslant 10^{7}\, \rm M_{\odot}$. The difference between AGN and DAGN distributions vanishes at lower redshift.
Second, in the highest-resolution TNG50, most of the DAGN (and AGN) are powered by seed-mass MBHs (i.e., with $\sim 10^{6}\, \rm M_{\odot}$ in that simulation). More evolved MBHs power the DAGN in the other simulations.

In any case, we find relatively small differences between the distributions of MBH masses for the full AGN populations and those for the DAGN populations.

The distributions of the AGN and DAGN host galaxy stellar masses shown in Fig.~\ref{fig:hist_Mstar} are also relatively similar (and with the same trends as identified for $M_{\rm MBH}$ for all simulations), although varying quite strongly from one simulation to another. For example, in EAGLE no AGN (and thus DAGN) are powered in low-mass galaxies with $M_{\star}\leqslant 10^{9.8}\, \rm M_{\odot}$. This is due to the combination of a relatively low seed mass, a modified Bondi accretion model including the effect of gas angular momentum \citep[][without a boost factor]{2015MNRAS.454.1038R}, and the fact that MBHs only enter an efficient growth phase when their haloes reach the critical halo virial temperature of $T_{\rm vir}\sim 10^{5.6}\, \rm K$ \citep[][]{2018MNRAS.481.3118M}. In SIMBA, gravitational torque accretion feeds low-mass MBHs efficiently \citep{2013ApJ...770....5A,2015ApJ...800..127A,2017MNRAS.472L.109A} but the low number of AGN and DAGN in low-mass galaxies is due to seeding taking place in galaxies of $10^{9.5}\, \rm M_{\odot}$ \citep[see discussion in][]{2019MNRAS.486.2827D,thomas2019black,2022MNRAS.514.4912H}.
All the other simulations produce a larger number of AGN and DAGN in galaxies with $M_{\star}\leqslant 10^{10}\, \rm M_{\odot}$ with median values of $\rm log_{10} M_{\star, \, DAGN}/M_{\odot}= 9.45, 9.74$ for Illustris and TNG100  at $z=4$ for example (top row in Fig.~\ref{fig:hist_Mstar}).
This is due to a relatively weak SN feedback in Illustris and a combination of a high seeding mass and weaker SN feedback towards lower redshifts in TNG, both effects allowing for efficient MBH accretion and the emergence of AGN. Interestingly, the lower resolution of the gas spatial distribution in TNG300 reduces MBH accretion rates and lowers the number of DAGN involving galaxies of $\sim 10^{9}\, \rm M_{\odot}$ compared to TNG50 and TNG100.

In contrast, from Fig.~\ref{fig:hist_Mstar}, we identify differences in the most massive galaxies hosting DAGN at $z=0$. In TNG50/TNG100 and SIMBA, no galaxies above $10^{10.2}\, \rm M_{\odot}$ power a DAGN. This is due to the efficient AGN feedback in larger galaxies in these simulations \citep[see details in][Fig.~2]{2021MNRAS.tmp.2905H}. A drop in the ability to power an AGN is found for more massive galaxies ($\sim 10^{11}\, \rm M_{\odot}$) in the other simulations like Illustris, or no drop at all in Horizon-AGN. The feedback of AGN themselves can play a role in the characteristics of the DAGN and in which galaxies they live, and given the differences in the feedback modelling employed in these simulations, there is no consensus here.

All simulations predict that DAGN tend to be slightly more massive (no more than 0.35 dex when considering all the simulations)
and be found in more massive galaxies than the overall AGN population at higher redshift (see Fig.~\ref{fig:hist_Mstar}). Primary MBHs of the pairs are, on average, hosted by more massive galaxies than the AGN population. 
There is a turnover at lower redshift, with DAGN being powered by slightly less massive MBHs (for example, at $z=0$) and located in galaxies less massive than the entire AGN population. The trend is driven by the secondary MBHs being embedded on average by less massive galaxies than the global AGN population.
We identify the largest difference in TNG50 and Illustris, where, for example, in the latter simulation AGN have a median of $M_{\rm MBH}= 10^{8}\, \rm M_{\odot}$ (in galaxies with a median of $M_{\star}= 10^{10.8}\, \rm M_{\odot}$) and DAGN of $M_{\rm MBH}= 10^{7.2}\, \rm M_{\odot}$ ($M_{\star}= 10^{10.4}\, \rm M_{\odot}$).  

\subsubsection{Are DAGN over-massive with respect to their hosts?}
In Fig.~\ref{fig:hist_Mbh_Mstar}, we interpret these results with the $M_{\rm MBH}/M_{\star}$ ratios, i.e. the over-massiveness of DAGN compared to the AGN population. 
Simulated DAGN tend to involve MBHs with $M_{\rm MBH}/M_{\star}$ ratios similar to those of the entire AGN population in $z\geqslant 2$, in all simulations. We note that DAGN in TNG50 involve slightly more massive MBHs with respect to their galaxies, compared to the full AGN population at $z=4$; the trend, however, vanishes at $z=2$.
We identified larger differences at $z=0$ in a few simulations. In Illustris, the DAGN are triggered by MBHs relatively less massive with respect to their host galaxies (median of $\log_{10}\, M_{\rm MBH}/M_{\star}=-3.25$) in comparison to the full AGN population (median of -2.81). Conversely, the TNG50 and Horizon-AGN simulations have a few more systems with $\log_{10}\, M_{\rm MBH}/M_{\star}\sim -1.5--2$ in the DAGN population compared to the AGN one. This leads to a median of $\log_{10}\, M_{\rm MBH}/M_{\star}=-2.24$ for DAGN and $-2.38$ for AGN in TNG50, and $-2.65$ for DAGN and $-2.84$ for AGN in Horizon-AGN.

We also study the over-massiveness of the primary and secondary MBHs of the pairs (see caption of Fig.~\ref{fig:hist_Mbh_Mstar}). At high redshift ($z>2$), all simulations predict that the primary MBHs are, on average, slightly over-massive with respect to their hosts than the secondary MBHs and compared to the global AGN population. However, at $z=0$ there is no such consensus among the simulations. For example, in TNG50, the primary MBHs have a smaller median than the AGN distribution, and the secondary MBHs have a higher median.

\subsection{Bolometric luminosity of dual AGN}
There is a consensus that in all the simulations (except EAGLE, and Horizon-AGN at high redshift), a significant fraction of the DAGN population (defined as $L_{\rm bol}\geqslant 10^{43}\, \rm erg/s$) is driven by bright AGN, brighter than the average full AGN population.   This agrees with the fact that the DAGN MBHs are more massive than the full AGN population at $z>0$.
The excess does not exceed more than half a dex, as indicated by the medians of the $L_{\rm bol}$ distributions of DAGN and AGN shown in Fig.~\ref{fig:hist_Lbol}.
In TNG50, for example, DAGN have a median of $\log_{10}\, L_{\rm bol}\rm /(erg/s)=43.9$ and AGN a median of $43.5$ at $z=2$.
We compute the Eddington ratios of DAGN and AGN and find slightly larger $f_{\rm Edd}$ for DAGN in the TNGs and EAGLE, slightly lower $f_{\rm Edd}$ in Horizon-AGN, and identical medians for the Illustris, SIMBA, and Astrid DAGN and AGN populations.

We find that the ``more luminous DAGN'' trend vanishes with time. Although with low number statistics, the median of the DAGN is similar to those of the AGN, with only a larger fraction of DAGN being triggered by AGN with $L_{\rm bol}\geqslant 10^{44}\, \rm erg/s$ in TNG50 and Horizon-AGN at $z=0$ (bottom panels).

There is a consensus among the simulations that the population of primary MBHs of the DAGN (vertical solid lines) are, on average, significantly brighter than the secondary MBHs (vertical dashed lines) and the AGN population (grey arrows).  
The difference between the median luminosity of the primary and secondary MBHs is highest at high redshift and decreases with time. However, when looking at the individual DAGN systems we find that a significant fraction have a secondary MBH brighter than the primary one, in all the simulations. At $z=4$, $0$ to $34\%$ DAGN include a brighter secondary MBHs (EAGLE or the lowest hand, and Horizon-AGN for the higher hand). About $18-66\%$ have brighter secondary MBHs at $z=2$ (SIMBA and EAGLE, respectively), and $23-55\%$ at $z=0$ (TNG100, TNG300, respectively).

Understanding the fueling of the MBHs due to the interactions triggered by the mergers can be done to some extend by investigating the luminosity of the primary and secondary MBHs as a function of their separation. The efficiency of the fueling depends on MBH mass, the mass of the hosts, and their mass ratios as interactions are different in the case of minor and major galaxy mergers. 
While the effect of small galaxies will remain small on the primary ones, the effect of the primary galaxies on the secondary ones can be important. In the late stage of the merger, the gas can be more efficiently funnelled into the central galactic region of the MBHs located in the secondary galaxies, as shown with high-resolution zoom-in simulations \citep[e.g.,][]{2017MNRAS.469.4437C}. This often results in secondary MBHs brighter than primary MBHs and even more so as the pairs shrink in size.
In the simulations, a significant fraction of DAGN have a secondary MBH (the least massive) brighter than the primary MBH. Yet, the luminosity ratios $L_{\rm bol, \, secondary}/ L_{\rm bol, \, primary}$ are close to unity, and vary in the range $0.90-1.10$ for all the simulations. The ratios do not significantly evolve with the separation of the DAGN. A major limitation with these simulations and their resolutions is that most DAGN have a separation larger than 5-10 pkpc, which prevents any investigation on fueling processes at smaller separation. For the present simulations, AGN feedback processes from the primary MBHs could prevent their accretion more significantly than the feeback from the secondary (less massive) MBHs, and explain the non-evolution of the ratios with DAGN separation.

\subsection{Are dual AGN powered by central and/or satellite galaxies?}
We pursue our comparison to understand what type of galaxies host the DAGN.
In this section, we quantify how many pairs are composed of central-central, satellite-satellite and central-satellite galaxies at redshift $z= 4, \, 2, \,0$. The results are shown in Table \ref{tab:sat_cen_pairs}. Central galaxies are the most massive systems in their respective haloes, whereas satellite galaxies refer to all the other systems, but the most massive in haloes.
We specify here that in this section, we rely on the identification of central and satellite galaxies made by the different groups which performed the simulations; we do not reprocess the simulations to extract this information in a uniform way. 

In short, we find no consensus in the cosmological simulations of the field on which galaxy systems power the highest number of DAGN. In addition, results vary quite significantly with resolution if we compare TNG50 with the lower-resolution TNG100 and TNG300:  at $z=4$ all TNG50 DAGN are central-central systems (low statistics with less than 10 AGN in these systems), while most of the TNG100 and TNG300 systems are composed of both satellite galaxies, all three simulations have the same behaviour at $z=2$ with most of the DAGN being hosted by central-central galaxies, while finally at $z=0$ they show again the opposite behaviour with TNG50 DAGN being all satellite galaxies (low statistics) and most of TNG100 and TNG300 DAGN being central-central systems. Given the low statistics in some categories of systems, our results for TNG50 may not represent what would be found in a larger volume with the same high resolution. 

At $z=2$ for which we have the highest statistics, we find that Illustris and TNG50/TNG100/TNG300 predict that more than $\sim$80$\%$ of the DAGN are in central-central galaxy systems while both EAGLE and SIMBA predict that the most common type ($\geqslant 80\%$) of DAGN are central-satellite galaxies. 

We also find that SIMBA predicts the highest fraction of satellite-satellite pairs ($20\%$) while the rest of the simulations present $0-2\%$ for this category. In SIMBA, because MBHs form in relatively massive galaxies with $M_{\star}\geqslant 10^{9.5}\, \rm M_{\odot}$ they need to grow efficiently to reach the empirical $M_{\rm BH}-M_{\star}$ scaling relation observed in the local Universe. As a result, a large fraction of galaxies with $M_{\star}\sim 10^{9.5}\, \rm M_{\odot}$ (of which many are satellite galaxies) host an AGN. Those are responsible for the high fraction of satellite galaxies involved in DAGN in SIMBA with respect to the other simulations. EAGLE predicts that all its DAGN are driven in central-satellite systems, but the low statistics (less than 10 AGN) prevents one from drawing any robust conclusions.

\begin{figure*}
    \centering
    \includegraphics[scale=0.52]{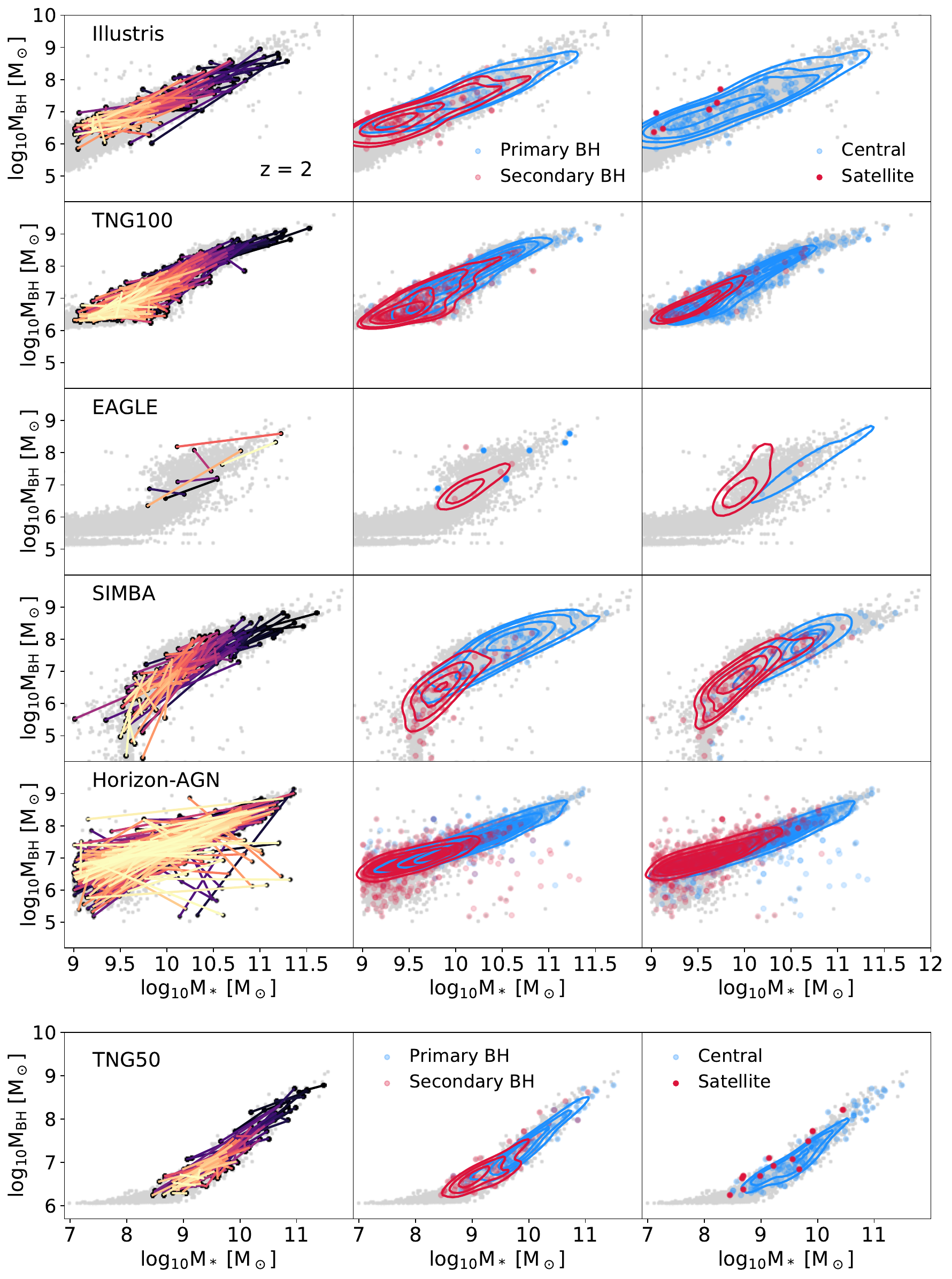}
    \caption{$M_{\rm MBH}-M_{\star}$ diagram at $z=2$. \textit{Left panels}: MBHs in DAGN are joined with coloured lines. \textit{Central panels}: All DAGN are shown as dots labelled as primary (most massive, \textit{blue}) or secondary (less massive, \textit{red}) MBHs with contour levels overlaid on top. \textit{Right panels}: As in the central panels, but distinguishing between MBHs embedded in central (\textit{blue}) or satellite (\textit{red}) galaxies.
    Only galaxies with $M_\star \geqslant 10^{9} \,\rm M_\odot$ ($M_\star \geqslant 10^{7} \,\rm M_\odot$ for TNG50) and AGN with $L_{\rm bol}\geqslant 10^{43}\, \rm erg/s$ and separated by $d\leqslant30\, \rm pkpc$ are considered. 
    }
    \label{fig:central_sat}
\end{figure*}

To illustrate our findings, we show in Fig.~\ref{fig:central_sat} the $M_{\rm MBH}-M_{\star}$ diagram of the simulations at $z=2$. All DAGN identified in each simulation ($M_{\star}\geqslant 10^{7}\, \rm M_{\odot}$ for TNG50, and $\geqslant 10^{9}\, \rm M_{\odot}$ for the others $L_{\rm bol}\geqslant 10^{43}\, \rm erg/s$, $d \leqslant30 \, \rm pkpc$) are joined with solid coloured lines in the left panels. To guide the eye, we also represent the full MBH population in the background of each panel with grey points. In the middle panels, we divide the DAGN into their primary and secondary MBHs, i.e. the most and least massive MBHs of the pairs (shown in blue and red, respectively).

Regarding the primary and secondary MBH distributions at $z=2$, all simulations seem to agree that the secondary MBH population is restricted to smaller MBH and galaxy stellar masses ($M_{\rm MBH}\leqslant 10^{8-8.5}\, \rm M_{\odot}$, $M_{\star}\leqslant 10^{10.5}\, \rm M_{\odot}$) than the primary MBHs, on average. 
On the other hand, the primary MBH population expands almost the whole stellar mass range ($M_{\star}= 10^{9-11.5}\, \rm M_\odot$), except for the lowest-mass MBHs and/or smallest galaxies. 
For example, in SIMBA, no more than one MBH with $M_{\rm MBH}\leqslant 10^{6}\, \rm M_{\odot}$ is a primary MBH in the DAGN systems, while a significant fraction of secondary MBHs are in this mass range.

Central and satellite galaxies are shown in blue and red symbols on the right panels. We identify two different behaviours in our sample of simulations.
In Illustris, TNG50, TNG100, EAGLE, MBHs identified as belonging to satellite galaxies have on average higher $M_{\rm MBH}$ for the same $M_\star$ with respect to those in central galaxies. Those could be largely driven by ongoing galaxy mergers, where the merging central galaxies have been flagged as satellites of the other central galaxies.
In SIMBA, the satellite and central galaxies have very similar distributions in the $M_{\rm MBH}-M_{\star}$ plane.\\

\iffalse
\begin{table*}
\caption{Study of different types of pairs for the DAGN identified in the different simulations at redshift $z\sim2$. \mh{It would be fantastic to add $z=0$, $z=1$ and $z=4$.}\mh{Alternatively, we could derive this for all redshifts and make a figure as a function of redshift.}}
\label{tab:sat_cen_pairs}
\begin{tabular}{lcccccc}
\hline
                             & \textbf{Illustris} & \textbf{TNG50} & \textbf{TNG100} & \textbf{TNG300} & \textbf{EAGLE} & \textbf{SIMBA} \\ \hline
Central-Central {[}\%{]}     & 93                 & 86             & 82              & 91              & 0              & 0              \\
Satellite-Satellite {[}\%{]} & 1                  & 2              & 2               & 0               & 0              & 20             \\
Central-Satellite {[}\%{]}   & 6                  & 12             & 16              & 9               & 100            & 80             \\ \hline
\end{tabular}
\end{table*}
\fi

\begin{table*}
\caption{Percentage ($\%$) of DAGN embedded in systems where both galaxies are central galaxies, both satellite galaxies and mixed systems. Only DAGN with $L_{\rm bol}\geqslant10^{43}\, \rm erg/s$ in galaxies with $M_{\star}\geqslant 10^{9}\, \rm M_{\odot}$ ($M_{\star}\geqslant 10^{9.5}\, \rm M_{\odot}$ for SIMBA) are considered. The percentages based on low statistics (less than 10 DAGN of a given type) are written in parentheses.
} 
\label{tab:sat_cen_pairs}
\begin{tabular}{llllllllll}
\hline
&  \multicolumn{3}{c}{Central-Central} & \multicolumn{3}{c}{Satellite-Satellite}  & \multicolumn{3}{c}{Central-Satellite}\\
\hline
Redshift &  \multicolumn{1}{|l}{$z=4$} & $2$ & $0$ & \multicolumn{1}{|l}{$z=4$} & $2$ & $0$ & \multicolumn{1}{|l}{$z=4$} & $2$ & $0$\\
\cline{2-10} 
& \multicolumn{1}{|l}{} & &&\multicolumn{1}{|l}{} &&&\multicolumn{1}{|l}{} &&\\
\textbf{Illustris}  & \multicolumn{1}{|l}{80} & 93 & (67) &\multicolumn{1}{|l}{0} & 1 & (0) &\multicolumn{1}{|l}{20}& 6 & (33)\\
\textbf{TNG50}      & \multicolumn{1}{|l}{(100)} & 86 & (0) &\multicolumn{1}{|l}{(0)}& 2 & (100) &\multicolumn{1}{|l}{(0)}& 12 & (0) \\
\textbf{TNG100}     & \multicolumn{1}{|l}{1.3} & 82 & 82 &\multicolumn{1}{|l}{85.3}& 2 & 5 &\multicolumn{1}{|l}{13.3}& 16 & 13 \\
\textbf{TNG300}     & \multicolumn{1}{|l}{1} & 87 & 78 &\multicolumn{1}{|l}{94}& 1 & 4 &\multicolumn{1}{|l}{5}& 12 & 18\\
\textbf{EAGLE}      & \multicolumn{1}{|l}{(0)} & (0) & - &\multicolumn{1}{|l}{(0)}& (0) & - &\multicolumn{1}{|l}{(100)}& (100) & - \\
\textbf{SIMBA}      & \multicolumn{1}{|l}{0} & 0 & (0) &\multicolumn{1}{|l}{9}& 20 & (33) &\multicolumn{1}{|l}{91}& 80 & (67)\\
\textbf{Horizon-AGN}      & \multicolumn{1}{|l}{63} & 8 & 0 & \multicolumn{1}{|l}{2}&  16 & 4 & \multicolumn{1}{|l}{35}& 76 & 96 \\
\hline
\end{tabular}

\end{table*}

In this Section~\ref{sec:galaxies}, we demonstrated an evolution with time of the properties of MBHs embedded in dual systems and those of their host galaxies. 
At redshifts higher than the peak of the DAGN number density, we find that, on average, DAGN are powered by slightly more massive MBHs, are more luminous, and are located in more massive galaxies than the average population of AGN. 
However, at $z=0$ we find that although DAGN are powered by more massive MBHs than at higher redshift (and located in more massive galaxies), they are less massive than the AGN population (and located in less massive galaxies than the average AGN hosts). There is no consensus on whether there are more luminous than the AGN in the simulations we study here.
At any redshift, DAGN's MBHs do not seem more massive with respect to their galaxies when compared to the full AGN population.
In addition to comparing the DAGN population with the full AGN population produced in each simulation, it is interesting to compare the DAGN populations (and AGN populations) across the simulations. Simulations of the field all produce different populations of MBHs and differences in their respective MBH and galaxy subgrid physics result in different AGN populations \citep{2021MNRAS.503.1940H,2021MNRAS.tmp.2905H,2020ApJ...895..102L}. We found no perfect agreement in the simulations on the mass of the MBHs that composed DAGN, and in which galaxies they are located. In particular, the ability of a given simulation to have low-mass MBHs in DAGN depends on its seeding mass, accretion model, and SN feedback. In some simulations, efficient AGN feedback prevents massive MBHs in galaxies with $> 10^{10}\, \rm M_{\odot}$ to form DAGN.

\section{Detectability of dual AGN}
\label{sec:detectability}

\begin{figure*}
    \centering
    \includegraphics[scale=0.52]{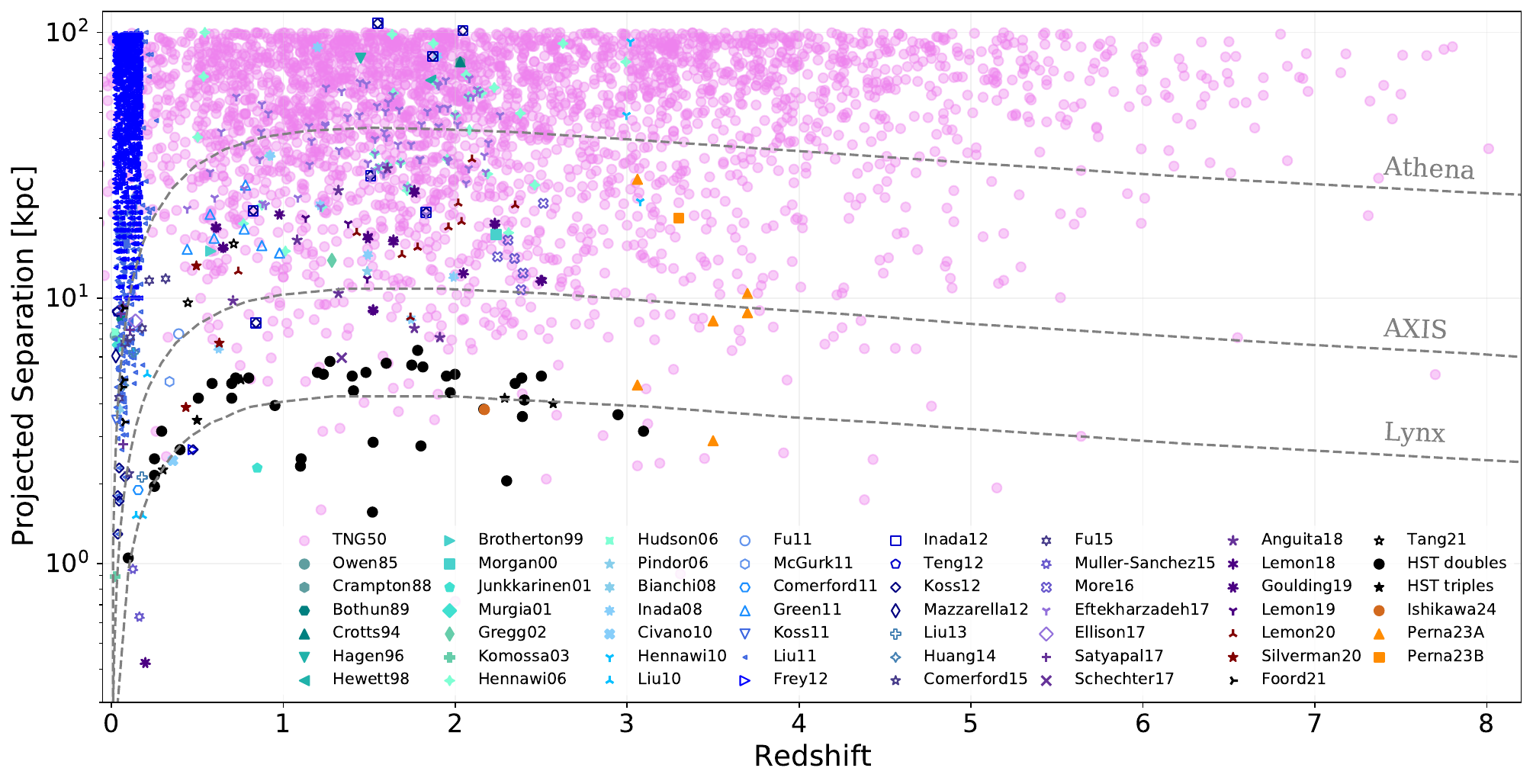}
    \caption{DAGN (and candidates) identified in observations up to $z\sim3$ (different luminosity thresholds have been used for these studies); adapted from 
    \citet{2022ApJ...925..162C}. Future X-ray telescopes will resolve separations of a hundred kpc down to a few kpc. Theoretical separations are shown with dashed lines for Athena, Lynx, and AXIS; the separation corresponding to eRosita is above Athena's line (outside the range of the figure). 
    We show the DAGN with kpc-scale separation produced in the TNG50 cosmological simulation with pink symbols. The DAGN are identified in snapshots spaced every 0.5 dex in redshift scale and randomly displaced in bins of 1 dex for visibility. DAGN are identified as AGN with $L_{\rm bol}\geqslant 10^{42}\, \rm erg/s$ in galaxies with $M_{\star}\geqslant 10^{7}\, \rm M_{\odot}$. AXIS and Lynx NASA concept observatories in X-ray are designed to detect fainter DAGN than $10^{43}\, \rm erg/s$ as discussed in Section~\ref{sec:detections}.
    }
    \label{fig:overview}
\end{figure*}

Several next-generation observatories will revolutionise our understanding of AGN activity before and during galaxy mergers by substantially increasing the number of observed DAGN to cosmic dawn.
High angular resolution is crucial to spatially resolve DAGN. We present in Fig.~\ref{fig:overview} an adaptation of the figure presented in \citet{2022ApJ...925..162C} of all current DAGN and candidates identified in observations. We also add the recent JWST discoveries \citep{2023arXiv231003067P,2023A&A...679A..89P,2024arXiv240308098I}. We add to our figure the angular resolution of Athena, AXIS, and Lynx as a function of redshift and the simulated population of TNG50 DAGN.

While eROSITA can theoretically resolve the X-ray emission of dual AGN with a separation below 100 kpc in the local Universe ($z\leqslant 0.3$, the angular resolution of $\sim 16''$ on-axis, and $\sim 28''$ FoV-average, corresponding to the top of Fig.~\ref{fig:overview}), none have been identified. 
The better sensitivity and angular resolution of Athena ($\sim 5''$ on-axis, $\sim 6''$ for the Wide-Field Imager's FoV) could allow detection of fainter systems with separations of 20 kpc in the local Universe, and 50-100 kpc in the redshift range $z=3-10$. The AXIS ($\sim 1.25''$ on-axis and $\sim 1.50''$ FoV-average) and Lynx ($\sim 0.5''$ FoV-average) NASA concept missions aim at unprecedented sub-arcsecond angular resolution to disentangle dual AGN with a separation down to a few kpc all the way to $z=10$. For example, AXIS will identify DAGN with enough statistics for population studies up to $z=3.5$ \citep[conservative redshift,][]{2023arXiv231107664F}, and additional systems at higher redshift.
In addition, the high spectroscopic capacities of some of these missions will allow us to identify binary systems that we do not tackle in this paper \citep[][for a review]{2023MNRAS.519.5149D}. 
On a shorter timescales, the Roman telescope with its near-IR wavelength large-sky coverage could probe AGN luminosities down to $L_{\rm bol}\geqslant 10^{42}\, \rm erg/s$ beyond $z\geqslant 1$ with an angular resolution of $\sim 0.13''$ \citep[][for two recent CSS white papers]{2023arXiv230614990H,2023arXiv230615527S}.

\subsection{Number of detections with future X-ray observatories}
\label{sec:detections}

\begin{figure*}
    \centering
    \includegraphics[scale=0.58]{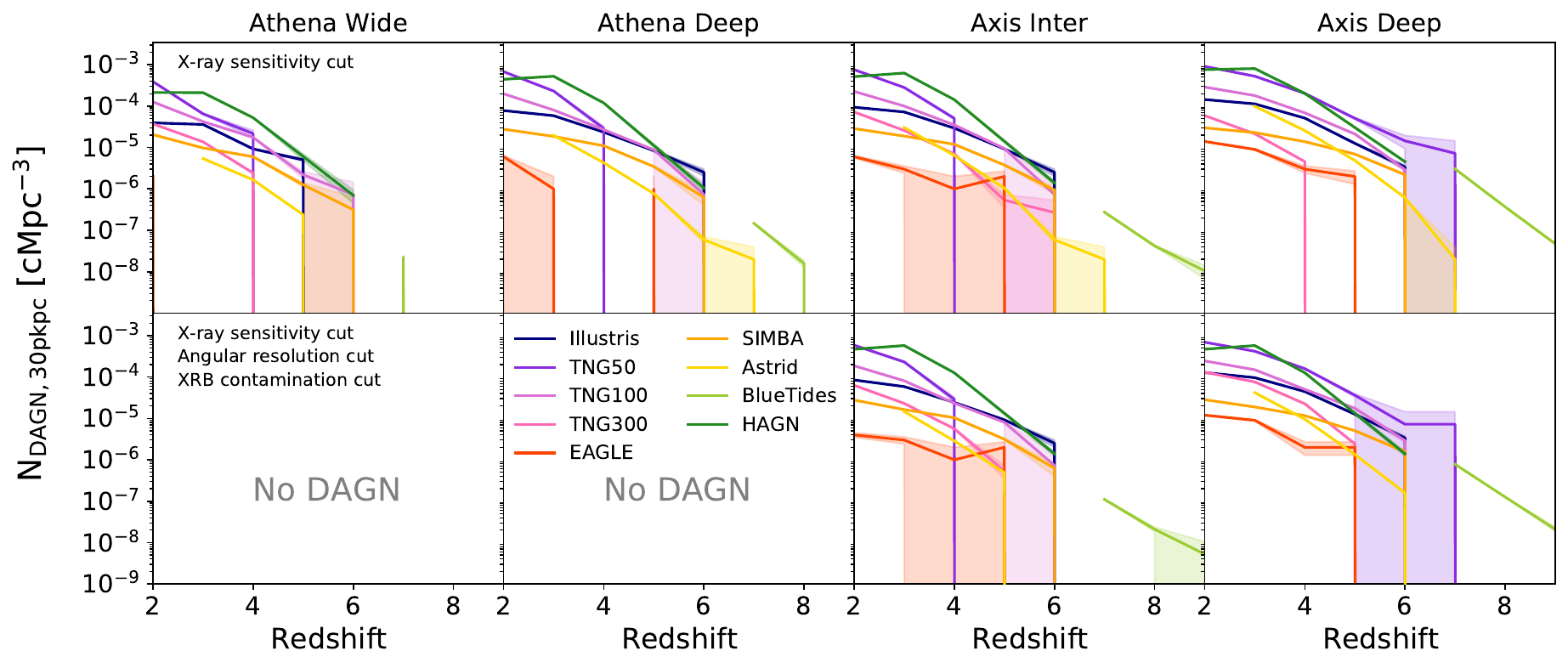}
    \caption{\textit{Upper panels}: Number density of DAGN as a function of redshift for next-generation X-ray telescopes (for the Wide and Deep surveys of Athena, and the Medium and Deep surveys of AXIS). Only the telescopes'  flux sensitivity is taken into account ($L_{\rm x}\geqslant L_{\rm thres}$, see a) of Section~\ref{sec:detections}). \textit{Lower panels}: Number density of DAGN above the flux sensitivity threshold, angular resolution, and galaxy-wide XRB emission ($L_{\rm x}\geqslant L_{\rm thres}, L_{\rm XRB}$).
    AGN are considered as dual if their 3-dimensional distance is $d \leqslant 30 $ pkpc (2-dimensional projected distances are discussed in the text).}
    \label{fig:futuremissions_new}
\end{figure*}

As most of nowadays' DAGN candidates are confirmed with X-ray observations, we focus on X-ray wavelengths in this section and predict the number of DAGN detectable by the future Athena and AXIS X-ray missions; detections with LynX are also discussed. To do so, we derive the luminosity of the AGN in the hard X-ray band (2-10 keV, see Eq.~\ref{eq:xray}) and consider the following aspects:\\

\noindent a) Sensitivity of telescopes. Only sources above the detection threshold of AXIS or Athena, $L_{\rm x}\geqslant L_{\rm thres}$, will be detectable. Here, we consider the highest sensitivity, i.e., corresponding to the $20\%$ of the surveys' total area. 
We use the two Athena surveys described in \citet{2013arXiv1306.2307N}. We refer to the first one as the Deep survey; it will cover an area of $5.28\, \rm deg^{2}$ and could have an exposure time of 1 Ms. The second survey, the Wide survey, is larger with an area of $47.42 \, \rm deg^{2}$ and an exposure of 90 ks. AGN with a luminosity in the range $L_{\rm x}=10^{42-43}\, \rm erg/s$ will be detectable up to $z=5$ in the Deep survey, and with $L_{\rm x}\geqslant 10^{43}\, \rm erg/s$ in the Wide survey for $z\geqslant 5$. At best capacity, the AGN with $L_{\rm x}\leqslant 10^{43}\, \rm erg/s$ will be detectable in the Deep survey up to $z=8$, and $> 10^{43}\, \rm erg/s$ beyond.
This paper employs the latest AXIS surveys described in \citet{2023arXiv231100780R,2023arXiv231107664F,2023arXiv231107669C}.
AXIS will cover $\rm 0.13 \, deg^2$ in its Deep survey with 5 Ms and detect AGN with $L_{\rm x}<10^{42}\, \rm erg/s$ for $z\leqslant7$, and $L_{\rm x}\sim 10^{42}\, \rm erg/s$ at $z=10$ for example. $2\rm \, deg^{2}$ will be covered in the AXIS Medium survey with 300 ks per pointing, allowing the detection of AGN with $L_{\rm x}\geqslant 10^{43}\rm \, erg/s$ up to $z=10$.  
\\

\noindent b)  We consider that only AGN with a stronger X-ray emission  than the galaxy-wide X-ray binaries emission from their host galaxies are detectable ($L_{\rm x}\geqslant L_{\rm XRB}$). We follow \citet{2021MNRAS.508.4816S} and use the empirical relation derived in \citet{2019ApJS..243....3L} to compute $L_{\rm XRB}$ (hard 2-10 keV band) as follows:
    \begin{equation}
    L_{\rm XRB} = 10^{29.15}(1+z)^2\, M_\star+10^{39.73}(1+z)^{1.3}\,\rm SFR,
    \label{eq:xrb}
\end{equation}
    with $M_\star$ ($\rm M_{\odot}$) being the total stellar mass of the galaxy, $z$ the redshift, and SFR ($\rm M_{\odot}/yr$) the star formation rate. Such scaling relations are derived from the nearby Universe ($\leqslant 50\, \rm Mpc$), and are extrapolated to higher redshifts. Because our study extends to very high redshifts, we discuss the impact of the redshift dependence factor ($(1+z)$ factor of Eq.~\ref{eq:xrb}) on our results.\\
    
\noindent c) Although there is evidence for an increasing fraction of obscured AGN (by gas and dust in the ISM and on the line-of-sight) towards high redshifts \citep[e.g.,][]{2022A&A...666A..17G}, we neglect the possible obscuration of the simulated AGN. Our predicted numbers of detections can thus be seen as upper limits.\\

\noindent d) Observations constrain the projected separation of sources in the sky, not their three-dimensional separation. In the following, we compare the number of DAGN using the 3D distance (as shown in Fig.~\ref{fig:futuremissions_new}) and the projected separations and consequences for future detections. We discuss the results when considering a distance of 20 cMpc for the projection of the pairs along the line of sight, which corresponds to an optimistic redshift uncertainty of $\Delta z \sim 0.05$ at $z\sim 6$, for example.\\

Fig.~\ref{fig:futuremissions_new} summarises the number density of detectable DAGN (3-dimensional separation of $\leqslant 30\, \rm pkpc$) for the Wide and Deep surveys of Athena and the Medium and Deep surveys of AXIS as predicted by the cosmological simulations of the field.
In the top panels, we first simply consider all DAGN above the sensitivity threshold of each survey ($L_{\rm x}\geqslant L_{\rm thres}$). As we take the sensitivity reached in the $20\%$ total survey, our predictions on the number of detections tend to be optimistic.
In the bottom panels, we remove the DAGN whose separation is below the angular resolution of the instruments and those located in star-forming galaxies whose X-ray emission from XRBs could contaminate and compromise the identification of the AGN.

\begin{figure*}
    \centering
    \includegraphics[scale=0.89]{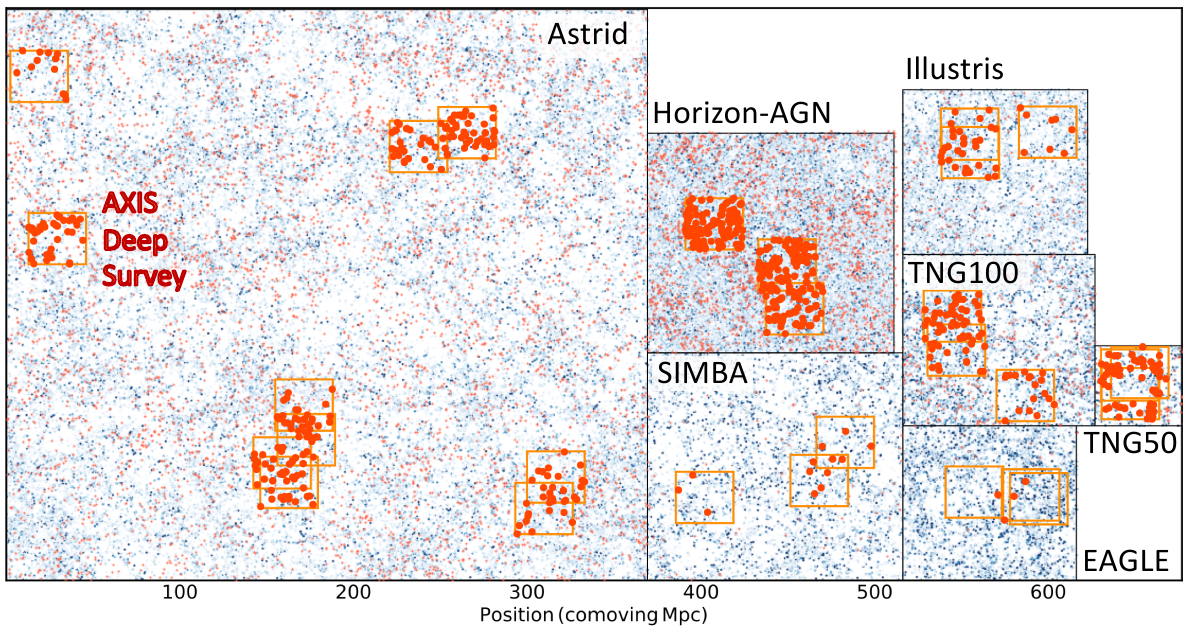}
    \caption{Illustration of the effect of the cosmic variance in detecting DAGN. Several random realisations of the AXIS Deep survey ($0.13 \, \rm deg^{2}$) are shown with orange squares. DAGN detections in these fields are shown with red points for the different simulations at $z=2$ ($z=3$ for Astrid). DAGN detections are defined as DAGN with luminosity above the AXIS sensitivity and angular resolution, and AGN brighter in X-rays than their galaxy-wide XRB population. Small blue dots represent all galaxies with $M_{\star}\geqslant 10^{9}\, \rm M_{\odot}$ (darker blue for more massive galaxies) in a slice of 100 cMpc (50 cMpc for TNG50), and small orange dots show all DAGN that are, in principle, detectable by the configuration of the AXIS Deep survey in the simulated volume.}
    \label{fig:map_detection}
\end{figure*}

We find that the angular resolution is the most limiting factor in detecting DAGN. For example, the lowest angular resolution of Athena would prevent the detection of systems with separation of $\leqslant 30 \, \rm pkpc$. Identifying DAGN separated by more than 30 pkpc is still relevant for understanding MBH growth and the connection to galaxy evolution. Based on sensitivity, these high-separation DAGN could start to be missed in the Athena Wide survey at $z\geqslant 5 / 3$ ($20\%$/rest of the survey) and at $z\geqslant 9 / 4$ ($20\%$/rest of the survey) in the Deep survey.

The higher angular resolution of AXIS will allow DAGN to be hunted with smaller separation of $d\leqslant 30\, \rm pkpc$ and at even higher redshifts. The detection of some DAGN with $L_{\rm x}\geqslant 10^{43}\, \rm erg/s$ could only be missed at $z\geqslant 10 / 7$ ($20\%$/ rest of the survey) in the Medium survey. If they exist, none of these bright DAGN would be missed in the AXIS Deep survey.

We find that the identification of DAGN with $L_{\rm x}\geqslant 10^{43}\, \rm erg/s$ is never hampered by the emission of XRBs. However, confirming the faintest AGN in the high-sensitivity surveys of AXIS could become challenging beyond the low-redshift Universe. 
First, if we consider the model of \citet{2019ApJS..243....3L} to compute galaxy-wide XRB emission without the redshift dependence, we find that only star-forming galaxies with $\rm \log_{10}\, SFR/(M_{\odot}/yr)\geqslant 1.5$ (independently of their stellar mass) would have an XRB hard X-ray luminosity reaching $10^{41}\, \rm erg/s$ \citep[see Fig.~2 in][]{2021MNRAS.508.4816S}, i.e. the best sensitivity achievable with AXIS.
Second, if we now consider the redshift dependence of the model, the XRB galaxy-wide luminosity increases with redshift.
At $z\geqslant 2$, the total XRB hard X-ray luminosity could reach the sensitivity of the AXIS Deep survey in galaxies with $M_{\star}\sim 10^{9}-10^{10}\, \rm M_{\odot}$ if they form stars with $\rm \log_{10}\, SFR/(M_{\odot}/yr)=0.5$, and potentially in all star-forming galaxies with $M_{\star}\geqslant 10^{11}\, \rm M_{\odot}$ (and $\rm \log_{10}\, SFR/(M_{\odot}/yr)=-3$).

In Fig.~\ref{fig:futuremissions_new}, we have shown the number density of DAGN identified with their 3-dimensional separation. Projection onto a redshift slice corresponding to $20\, \rm cMpc$ can artificially increase the number of detections by a factor of 2 at $z=2$ or $z=4$ for the AXIS Medium survey and up to 2.4 for the Deep survey.

Here, we do not explicitly show the results for the Lynx observatory. Still, the higher sensitivity and better angular resolution would potentially allow for the detection of more DAGN while being more limited by the small size of the surveys (see next section). Paradoxically, the theoretical gain of the higher sensitivity may, in reality, be diminished by a higher XRB contamination that would result in a challenging confirmation of the presence of AGN in these X-ray observations. Such an observatory is, however, needed to investigate DAGN when evolving in the same galaxies.

Based on what we discuss in this section, the Deep survey of AXIS will provide us with the best chance of observing AGN pairs at any redshift with expected number densities ranging from $10^{-5}-10^{-3}\, \rm cMpc^{-3}$ at $z=2-3$ to $10^{-6}-10^{-5}\, \rm cMpc^{-3}$ at $z\sim 6$ across the simulations.
In the next section, we derive the number of expected DAGN detections for the actual size of the AXIS surveys and investigate whether cosmic variance will challenge the detections.

\subsection{Is cosmic variance a limiting factor to identify dual AGN?}
The AXIS Deep survey will have a small size of $0.13\, \rm deg^{2}$ to obtain an unprecedented sensitivity. As such, it is fundamental to study the impact of cosmic variance on the number of detections. To do so, we randomly select in the simulations' volume 100 AXIS Deep surveys with a redshift slice of 100 cMpc each (i.e., a slice with a depth of 100 cMpc in the z-direction of a simulation box), and we derive the number of detections for each of them (i.e. DAGN identified with all criteria mentioned in the previous section).
At $z=2$, AXIS would detect a mean of  
$16.5^{22}_{11}$, $47.4^{54}_{39}$, $32.1^{45}_{11}$, $101.3^{139.1}_{74.9}$,$2.1^{4}_{0}$, $4.1^{6}_{2}$ systems
for Illustris, TNG50, TNG100, Horizon-AGN, EAGLE, SIMBA (upper and lower values represent the 15 and $85\%$ percentiles). For Astrid's last $z=3$ output, we find that the Deep survey could detect $25.3^{37}_{13}$ in a small redshift slice of 100 cMpc. A visualization of several surveys for all the simulations is shown in Fig.~\ref{fig:map_detection}. The higher number of DAGN in the Horizon-AGN simulation partially results from a higher number of galaxies as the galaxy mass distribution of Horizon-AGN has a higher normalization than the other simulations studies here.

These lower limits have to be cumulated with the probability of finding DAGN at all redshifts, i.e., in the observed light cone. As our analysis does not consider DAGN within single galaxies, a mission like AXIS with high angular resolution could identify more DAGN.

The large volume of Astrid also allows us to investigate the cosmic variance expected for the AXIS Medium survey. At $z=3$ (redshift slice of 100 cMpc), a mean of $117^{168}_{57}$ DAGN are detectable per survey (averaged over 100 realizations). With our field realizations, we find that a minimum of 35 and a maximum of 217 systems could be identified, as these numbers are lower limits on the detections; this is very encouraging. 
At higher redshift, DAGN are less numerous and which will complicate their detections. Nevertheless, at $z=5$ about $2.81^{5.1}_{1}$ and $8.61^{14.1}_{3}$ DAGN could be detected in the Deep and Medium surveys, respectively. 
A mission such as AXIS will have the power to detect and characterise a large number of multiple AGN, and likely a few all the way up to the end of the reionization era.

\subsection{Are dual AGN precursors of the LISA gravitational wave antenna detections?}
\begin{figure}
    \centering
    \includegraphics[scale=0.475]{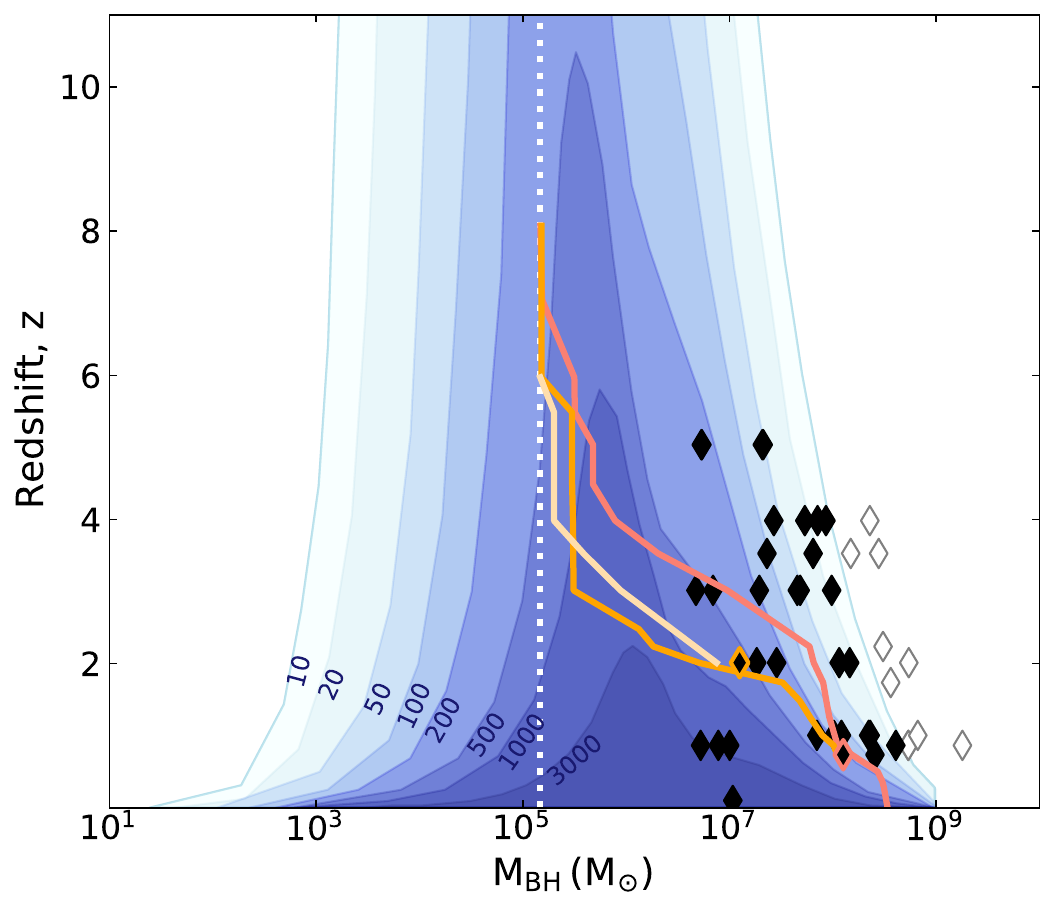}
    \caption{Growth history of 3 EAGLE MBHs identified as two DAGN (salmon, orange, and white lines). The seeding mass used in EAGLE is shown as a vertical white dotted line at $M_{\rm MBH}\sim 10^{5}\, \rm M_{
    \odot}$.
    The mergers of these two DAGN are shown with black diamonds with edge colours ($z=2$ and $z=0.86$).
    Mergers of the other DAGN are shown with black diamonds if detectable by LISA and white diamonds if they are outside LISA signal-to-noise ratios (shown in blue).
    }
    \label{fig:LISAdet_EAGLE}
\end{figure}

To have a first glance at whether the DAGN that we identified in the large-scale cosmological simulations (with a 3-dimensional separation of $\leqslant 30\, \rm pkpc$ and $L_{\rm bol}\geqslant 10^{43}\, \rm erg/s$) could merge and be detected by LISA, we follow the merger trees of their host galaxies in time. For simplicity, we restrict ourselves to the EAGLE simulation. The simulation does not model MBH dynamics, and we do not model any time delays to decouple MBH mergers from those of their host galaxies. 
Delays have been included post-processing in some studies and showed to significantly shift in time the coalescence of MBHs \citep[e.g.,][]{2016MNRAS.463..870S,2017MNRAS.471.4508K,2020MNRAS.491.2301K, 2020MNRAS.498.2219V}.
In Fig.~\ref{fig:LISAdet_EAGLE}, we show the history of 2 pairs of dual MBHs shown by the yellow and orange line and the orange and red line further in time. The merger of these dual MBHs is indicated by black-filled symbols outlined in orange. All the other black-filled symbols are mergers of all other DAGN in the simulation.
As in reality, we expect MBH coalescence to be decoupled from galaxy mergers \citep[][for a review]{2023LRR....26....2A}, we expect the mergers shown in Fig.~\ref{fig:LISAdet_EAGLE} (black symbols) to be shifted towards lower redshift and likely slightly larger MBH masses. Nevertheless, we find that a large fraction ($75\%$) of the DAGN that we identified in EAGLE with $L_{\rm bol}\geqslant 10^{43}\, \rm erg/s$ could merge within the LISA signal-to-noise region in  Fig.~\ref{fig:LISAdet_EAGLE} and be precursors of LISA gravitational wave events. The fraction is similar to what is found in Horizon-AGN without delays \citep{2022MNRAS.514..640V}; this number drops to $40\%$ when including post-processing delays to that simulation.

\section{Conclusions}
In this work, we quantified the number of dual AGN produced by the large-scale cosmological simulations of the literature in the redshift range $z=0-10$ (Illustris, TNG50, TNG100, TNG300, Horizon-AGN, EAGLE, SIMBA, BlueTides, and Astrid). We defined dual AGN as two MBHs with $L_{\rm bol}\geqslant 10^{43}\, \rm erg/s$ located in two distinct galaxies and separated by $\leqslant 30\, \rm pkpc$. With our definition, we do not study small-scale dual AGN of a few kpc or below, which are hosted by single galaxies. Thus, we restrict ourselves to systems currently observable in surveys (e.g., with JWST) and those which will be targeted, for example, by next-generation X-ray telescopes probing up to the high-redshift Universe. In other words, we investigate the possible precursors of LISA events that electromagnetic telescopes could detect. We summarise our main findings below. We did not consider the obscuration of the AGN by gas and dust; our numbers of dual AGN could thus be upper limits.

\begin{itemize}
    \item The number density of the dual AGN varies from $10^{-8}$ (or none) to $10^{-3}\, \rm cMpc^{-3}$ between the simulations and in the redshift range $z=0-7$ (Fig~\ref{fig:ndensity_fraction_z}, Table~\ref{tab:dual_AGN_30}). No dual AGN is identified at higher redshift, except in the largest-volume BlueTides simulation. 

    \item We find that the largest discrepancy in the simulations' number density of dual AGN on average takes place at $z=1-3$ (2 dex difference among the simulations, Fig~\ref{fig:ndensity_fraction_z}).
        
    \item Although most identifications of dual AGN (and candidates) are at $z\leqslant 1$ in observations (Fig.~\ref{fig:overview}), there is a consensus in the simulations studied here that the highest number of detections could be expected at cosmic high noon (i.e., $z=1-3$). The number of dual AGN thus peaks at about the same time (slightly higher redshift) as the AGN number density and cosmic star formation rate (Fig.~\ref{fig:DAGN_follow_AGN}).

    \item The fraction of dual AGN (defined as \#DAGN/\#AGN) depends on not only the population of dual AGN identified in the simulations but also the AGN populations themselves, which strongly depends on the subgrid physics implemented in each simulation (Fig.~\ref{fig:map}).
    In galaxies with $M_{\star}\geqslant 10^{9}\, \rm M_{\odot}$, we find dual AGN fractions of
    0-0.01 $\%$, 0-0.03 $\%$, 0.02-0.06 $\%$, 0.01-0.05 $\%$
    %0-0.8 $\%$, 0.6-2.8 $\%$, 1.9-5.0 $\%$, 1.4-4.6 $\%$
    in the local Universe, at $z=1$, $z=2$, $z=5$, respectively. The fractions peak in the range $z=2-4$ on average for most of the simulations studied here (Fig~\ref{fig:ndensity_fraction_z}).

    \item Our comparison with current observational constraints \citep{2011ApJ...737..101L,2012ApJ...746L..22K} indicates that all simulations could produce too few dual AGN at $z=0$ (Fig.~\ref{fig:comp_obs}). We also compared our results to several recent constraints at high redshift \citep[$z\sim 3$,][Fig.~\ref{fig:comp_obs}, Fig.~\ref{fig:silverman2020}]{2020ApJ...899..154S,2023arXiv231003067P,2023arXiv231202311S}. We find that although most of the simulations agree with the upper limit of \citet{2023arXiv231202311S}, some simulations could produce too many dual AGN at $z\sim 3$ and even more so if we include dual AGN associated with low-mass galaxies ($M_{\star}
    \geqslant 10^{7}\, \rm M_{\odot}$). The above finding has to be taken with care, as we have neglected the possible obscuration of the AGN.

    \item We characterise the properties of the dual AGN and their host galaxies with respect to the properties of the total AGN population in each simulation (Fig.~\ref{fig:hist_Mbh}, \ref{fig:hist_Mstar}, \ref{fig:hist_Mbh_Mstar}, \ref{fig:hist_Lbol}, \ref{fig:central_sat}). At high redshift ($z>2$) dual AGN with $L_{\rm bol}\geqslant 10^{43}
    \, \rm erg/s$ are generally powered by MBHs of $M_{\rm MBH}\sim 10^{5}-10^{9}\, \rm M_{\odot}$. Dual AGN are slightly more massive, brighter, and located in more massive galaxies than the average population of AGN. 
    By $z=0$, dual AGN are powered by a narrower range of MBHs than at higher redshift ($M_{\rm MBH}\sim 10^{6}-10^{8}\, \rm M_{\odot}$). We find that although they are, on average, powered by more massive MBHs (and located in more massive galaxies) than at high redshift, they are paradoxically, on average, less massive than the AGN population (and located in less massive galaxies than the average AGN hosts).

    \item The above differences are caused by the different MBH and galaxy subgrid physics employed in the simulations (Fig.~\ref{fig:map}). The ability of a given simulation to have low-mass MBHs powering DAGN depends on its seeding mass, accretion model, and SN feedback. Some simulations predict that a large fraction of DAGN are located in $M_{\star}\leqslant 10^{10}\, \rm M_{\odot}$. At the same time, in current observations across cosmic times, AGN seem to be located in more massive galaxies (when galaxy mass is known). This could highlight a possible disagreement with observations and imply that seeding masses are too high, accretion modelling too efficient, and/or feedback from SN or AGN not efficient enough in preventing accretion into MBHs. For that low-mass galaxy regime, JWST already provides us with extremely valuable constraints.  
    Conversely, efficient AGN feedback in some simulations prevents massive MBHs in galaxies with $> 10^{10}\, \rm M_{\odot}$ from forming DAGN.

    \item Dual AGN with $L_{\rm bol}\geqslant 10^{43}
    \, \rm erg/s$ seem more luminous than the AGN in the same luminosity range, although there is no complete consensus among the simulations (Fig.~\ref{fig:hist_Lbol}).

    \item We find that the primary MBHs (the most massive of the pairs) of the dual AGN are, on average, brighter and located in more massive galaxies than the secondary MBHs (the less massive ones), although with a non-negligible fraction of brighter secondary MBHs. There is less consensus on the over-massiveness of the primary and secondary MBHs with respect to their host galaxies. While at high redshift, primary MBHs are slightly more over-massive than the secondary, the trend diminishes with time and by $z=0$, some secondary MBHs are even more over-massive than the primary MBHs (e.g. in TNG50).

    Although simulations show differences in the distributions of dual AGN separations, we note that when there are enough statistics, the distributions are, on average, relatively flat (at $z\geqslant 2$, Fig.~\ref{fig:hist_dual_d_60}). The median separations are in the range $10-20 \, \rm pkpc$. On average, dual AGN tend to be more separated with decreasing redshift, for example, from $z=4$ to $z=2$ (at a later time, our analysis suffers from low statistics on dual AGN).

    \item Dual AGN with $L_{\rm x,\, 2-10\, keV}\geqslant 10^{43}
    \, \rm erg/s$ will only be detectable in X-ray in a large redshift range with Athena if they are separated by more than 30 pkpc. However, the higher angular resolution of AXIS could allow identifying a large number of dual AGN with $d\leqslant 30 \, \rm pkpc$ present in the volume of its Medium (mean of more than 100 detections in Astrid) and Deep surveys (from a few to 100 detections depending on the simulations, Fig.~\ref{fig:futuremissions_new}, \ref{fig:map_detection}). The better sensitivity would also allow us to search for dual AGN fainter than $10^{43}\, \rm erg/s$ up to high redshift. 
    Only the faintest detectable dual AGN in the surveys could suffer from difficult confirmation due to galaxy-wide X-ray emission from X-ray binaries, in particular in star-forming galaxies with $M_{\star}=10^{9-10}\,\rm M_{\odot}$ and $\log_{10}\, \rm SFR/(M_{\odot}/yr)\geqslant 0.5$ beyond $z=2$ and more massive galaxies independently of their SFR. 
\end{itemize}

\section*{Acknowledgment}
CPS acknowledges support from the Max-Planck-Institut für Astronomie and the Galaxy \& Cosmology Department. MH acknowledges support from Max-Planck Institute through the MPIA fellowship.  
DAA acknowledges support by NSF grant AST-2108944, NASA grant ATP23-0156, STScI grants JWST-GO-01712.009-A and JWST-AR-04357.001-A, Simons Foundation Award CCA-1018464, and Cottrell Scholar Award CS-CSA-2023-028 by the Research Corporation for Science Advancement, CXO grant TM2-23006X, JWST grant GO-01712.009-A, Simons Foundation Award CCA-1018464, and Cottrell Scholar Award CS-CSA-2023-028 by the Research Corporation for Science Advancement.
YRG acknowledges the support of the “Juan de la Cierva Incorporation” fellowship (ĲC2019-041131-I) 

\section*{Data availability statement}
The data from the Illustris and the TNG simulations can be found on their respective websites: https://www.illustris-project.org, https://www.tng-project.org.  The data from the EAGLE simulation can be obtained upon request to the EAGLE team at their website: http://icc.dur.ac.uk/Eagle/. The data from the SIMBA simulation can be found on the website: http://simba.roe.ac.uk/. The Horizon-AGN simulation is not public, but some catalogs are available at: https://www.horizon-simulation.org/data.html. Data from the BlueTides and Astrid simulations are available at https://bluetides-portal.psc.edu/simulation/1/ and https://astrid-portal.psc.edu/simulation/1/, respectively.
Catalogs including galaxy and MBH properties of the DAGN can be shared upon request for some simulations.

\bibliographystyle{mn2e}
\bibliography{reference,biblio_complete_faintQSOs,biblio_complete,biblio_complete-2}

\appendix
\setcounter{table}{0}
\renewcommand{\thetable}{A\arabic{table}}

\begin{table*}
\caption[Quantitative study of dual AGNs for Illustris, TNG50, TNG100, TNG300, EAGLE and SIMBA.]{Quantitative study of dual AGNs for the simulations: Illustris, TNG50, TNG100, TNG300, EAGLE, SIMBA, Astrid, BlueTides and Horizon-AGN. Both number of systems and their corresponding number density are provided. MBHs are considered
as AGN if they have $L_{\rm bol} \geq  10^{43}$ erg/s and as part of dual AGN if the distance between them is $d < 30$ pkpc.  ``Single AGN'' refers to dual MBHs composed of only one AGN. ``Non-active MBH pairs'' represent MBH pairs where both MBHs have luminosity below $10^{43}\, \rm erg/s$.
Only galaxies with $M_\star \geq 10^{9}$ $\rm M_\odot$ ($M_\star \geq 10^{7}$ $\rm M_\odot$ for TNG50) are considered for this study.  }
\label{tab:dual_AGN_30}
\resizebox{\textwidth}{!}{\begin{tabular}{llccccccccc}
\hline
                               &                                 & \textbf{Illustris} & \textbf{TNG50} & \textbf{TNG100} & \textbf{TNG300} & \textbf{EAGLE} & \textbf{SIMBA}
                               & \textbf{Astrid}
                               & \textbf{Horizon-AGN}
                               \\ \hline
\multicolumn{2}{l}{$\mathbf{z \sim 4}$}                                      &           &       & &        &          & & &   \\
\multicolumn{2}{l}{No. galaxies with MBHs}                        &   4.465  & 64.065   &  4.305   &      65.899  & 1.258  &  12.248  & 1.710.392 &  11.684 \\
\multicolumn{2}{l}{No. MBH pairs}                                 &      65   & 4 &    94   &    1.125   &  22   &   45  & 841 &  720 \\
\multicolumn{2}{l}{No. detectable dual systems}                   &     &      &        &        &       &       & &   \\
                               & Dual AGN  $ L_{\rm bol} \geq 10^{43} \rm erg/s$   &     56   & 3 &   75  &    406    &    3 &  43  & 841 & 536 \\
                               & Single AGN                    &     9 & 0   &    15   &     451  &    10 & 2   & 0 &  166 \\
                               & Non-active MBH pairs                           &     0  & 1  &    4  &      268    &    9  &  0  & 0 &  18 \\
                               \\
\multicolumn{2}{l}{No. density MBH pairs ($10^{-6}$ $\rm cMpc^{-3}$)}      &      55   & 29 &  69   &    41  &  22  &   14  & 17 & 252 \\
\multicolumn{2}{l}{No. density detectable dual systems ($10^{-6}$ $\rm cMpc^{-3}$)} &    &       &        &        &       &     & &     \\
                               & Dual AGN $ L_{\rm bol} \geq 10^{43}\, \rm  erg/s$  &     47 & 22    &   55 &   15   &    3 &   14  & 17 &  187 \\
                               & Single AGN                    &     8 & 0   &   11   &     16  &  10 &   1   & 0 &   58 \\
                               & Non-active MBH pairs                           &     0  & 7  &   3   &    10   &  9   &   0   & 0 &   6 \\ \hline
\multicolumn{2}{l}{$\mathbf{z \sim 2}$}                         &             &           &        &        &       &    & &      \\
\multicolumn{2}{l}{No. galaxies with MBHs}                        &    19.276 & 61.249     &   16.156   &    230.151   &  4.696  &  20.936  & - & 53.285 \\
\multicolumn{2}{l}{No. MBH pairs}                                 &     206  & 62    &   404   &    1.889  &   143  &   108   & - &   2.639 \\
\multicolumn{2}{l}{No. detectable dual systems}                   &      &     &        &        &       &     & &     \\
                               & Dual AGN  $ L_{\rm bol} \geq 10^{43}\, \rm  erg/s$  &     91  & 58    &   270    &    1.633   &   6   &   88  & -  & 1.261 \\
                               & Single AGN                    &      83  & 4   &    117   &    256   &   34   &  18   & - &  1.094 \\
                               & Non-active MBH pairs                           &      32 & 0   &    17  &   0    &    103  &   2 & - &  284 \\
                               \\
\multicolumn{2}{l}{No. density of MBH pairs ($10^{-6}$ $\rm cMpc^{-3}$)}      &     173 & 450    &  298   &    68  &  143   &    34  & - &   923 \\
\multicolumn{2}{l}{No. density detectable dual systems ($10^{-6}$ $\rm cMpc^{-3}$)} &     &      &        &        &       &     & &     \\
                               & Dual AGN   $ L_{\rm bol} \geq 10^{43}\rm\,  erg/s$  &      77  & 421  &  199   &  59    &   6  &   28  & - &  441 \\
                               & Single AGN                         &     70   & 29  &   86    &   9  &   34  &   6  & - & 383 \\
                               & Non-active MBH pairs                            &     27   & 0  &    13   &   0   &   103   &   1  & - &  99 \\ \hline
$\mathbf{z \sim 0}$                   &     &                                 &           &        &        &       &    & &     \\
\multicolumn{2}{l}{No. galaxies with MBHs}                        &     37.389  & 34.322    &  25.427  &    328.420    &   6.933 &  44.830  & - & 33.044 \\
\multicolumn{2}{l}{No. MBH pairs}                                 &      58 & 25    &   118  &   208   &    36 &    14  & - &  190 \\
\multicolumn{2}{l}{No. detectable dual systems}                   &     &      &        &        &       &      & &    \\
                               & Dual AGN $ L_{\rm bol} \geq 10^{43}\rm \, erg/s$  &     3 & 2     &  39    &  173  &  0  &    3   & - &  45 \\
                               & Single AGN                    &     15  & 14    &   55   &   35   &   3  &   5 & - &  82  \\
                               & Non-active MBH pairs                       &     40  & 9  &   24   &   0     &   33  &    6  & - & 63 \\
                               \\
\multicolumn{2}{l}{No. density of MBH pairs ($10^{-6}$ $\rm cMpc^{-3}$)}      &    49   & 182   &  87   &    8   &  36   &   4  & - & 66 \\
\multicolumn{2}{l}{No. density detectable dual systems ($10^{-6}$ $\rm cMpc^{-3}$)}     &            &           &        &        &       &    & &      \\
                               & Dual AGN $ L_{\rm bol} \geq 10^{43} \rm \, erg/s$ &     3  & 15    &   29  &  6   &    0   &  1  & - &  16 \\
                               & Single AGN                    &     13  & 102    &   41    &    1   &    3  &  2   & - &  29 \\
                               & Non-active MBH pairs                            &     34  & 65    &   18   &   0    &    33   &  2   & - & 22  \\ \hline
\end{tabular}}
\end{table*}

\newpage

\section{QUANTITATIVE Summary OF DUAL AGN IN SIMULATIONS}
We report in Table \ref{tab:dual_AGN_30} a summary of the quantitative properties of the simulations studied for $z=4, \,2,\, 0 $: the total number of galaxies with MBHs, the number of systems (in absolute and number density) where systems include MBH pairs with $d < 30$ pkpc, AGN pairs (i.e., DAGN) and pairs with only a single AGN (often referred to as offset AGN). AGN are defined as $L_{\rm bol} \geq 10^{43}$ erg/s, and non-active MBH pairs as the pairs in which both MBHs have $L_{\rm bol} < 10^{43}$.

\section{DAGN with a more restrictive luminosity cut}
\begin{figure*}
    \centering
    \includegraphics[scale=0.58]{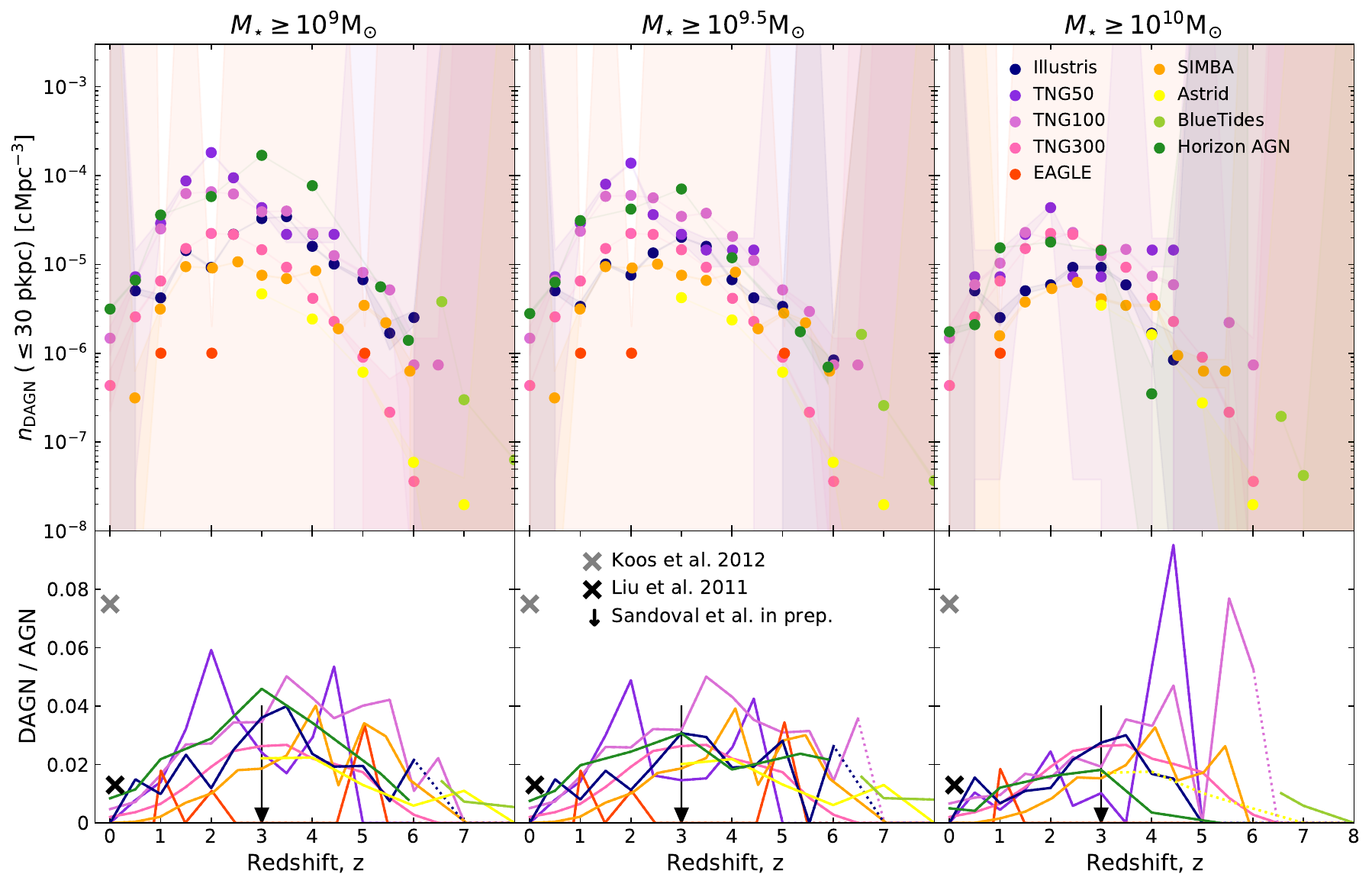}
    \caption{Same as Fig.~\ref{fig:ndensity_fraction_z} with an AGN luminosity limit of $L_{\rm bol}\geqslant 10^{44}\, \rm erg/s$ instead of $\geqslant 10^{43}\, \rm erg/s$. 
    }
\label{fig:ndensity_fraction_z_44}
\end{figure*}

In order to match better some current observational samples analyzing DAGN at high redshift \citep[e.g.,][]{2020ApJ...899..154S}, we show in Fig.~\ref{fig:ndensity_fraction_z_44} the analog of Fig~\ref{fig:ndensity_fraction_z} for a more restrictive cut in bolometric luminosity: $L_{\rm bol}\geqslant 10^{44}\, \rm erg/s$ instead of $10^{43}\, \rm erg/s$.

\label{lastpage}

\end{document}